\pdfoutput=1
\documentclass[a4paper,11pt]{article}

\usepackage{lineno}
\usepackage{jheppub}
\usepackage[utf8]{inputenc}
\usepackage{amsmath}
\usepackage{amsthm}
\usepackage{amssymb}
\usepackage{enumitem}   
\usepackage[english]{babel}
\usepackage{url}
\usepackage{mathtools}
\usepackage{bbold}
\usepackage{slashed}
\usepackage{multirow}
\usepackage{lipsum}
\usepackage{xcolor}
\usepackage{float}
\usepackage{revsymb4-2}
\usepackage{braket}
\usepackage{siunitx}
\usepackage{tensor}
\usepackage{xfrac}

\allowdisplaybreaks

\DeclareSIUnit\parsec{pc}

\newcommand{\bm}[1]{\boldsymbol{#1}}

\newcommand{\MP}{M_{\rm P}}

\newcommand{\D}{\mathrm{d}}
\newcommand{\p}{\partial}

\newcommand{\rot}{\mathrm{rot}}

\newcommand{\diver}{\mathrm{div}}

\definecolor{dgreen}{rgb}{0,0.6,0.0}

\newcommand{\FAISM}{FAI$_{\rm SM}$}

\begin{document}


\preprint{MS-TP-25-27}
\title{\Large Gravitational waves from axion inflation in the gradient expansion formalism. Part II. Fermionic axion inflation}

\author[a]{Richard~von~Eckardstein,}
\author[a, b]{Kai~Schmitz,}
\author[a, c]{and Oleksandr~Sobol}

\affiliation[a]{Institute for Theoretical Physics, University of M\"unster,\\
Wilhelm-Klemm-Stra{\ss}e 9, 48149 M\"{u}nster, Germany}

\affiliation[b]{Kavli IPMU (WPI), UTIAS, The University of Tokyo,\\
5-1-5 Kashiwanoha, Kashiwa, Chiba 277-8583, Japan}

\affiliation[c]{Physics Faculty, Taras Shevchenko National University of Kyiv,\\
64/13, Volodymyrska Street, 01601 Kyiv, Ukraine}

\emailAdd{richard.voneckardstein@uni-muenster.de}
\emailAdd{kai.schmitz@uni-muenster.de}
\emailAdd{oleksandr.sobol@uni-muenster.de}


\abstract{
    Axion inflation represents an intriguing source of gravitational waves (GWs) from the early Universe. In a companion paper~\cite{vonEckardstein_PAI_2025}, we previously leveraged the gradient expansion formalism (GEF) to investigate pure axion inflation (PAI), i.e., axion inflaton coupled to a pure gauge sector. In this paper, we extend our analysis to fermionic axion inflation (FAI), i.e., we allow for the presence of fermions in the gauge sector. PAI predicts a strongly blue-tilted GW spectrum; in our GEF benchmark study, all parameter regions leading to observable GWs turned out to violate the upper limit on the number of extra relativistic degrees of freedom, $\Delta N_{\rm eff}$. As we demonstrate in this paper, the situation is different for FAI: Schwinger pair creation of the charged fermions results in a damping of gauge-field production, which attenuates the GW signal. As a result, the GW signal from FAI can fall into the sensitivity reach of LISA and ET without violating the upper limit on $\Delta N_{\rm eff}$. This result notably applies to the arguably most realistic variant of Abelian axion inflation, in which the axion couples to the hypercharge sector of the Standard Model. Besides, we discuss GW emission from the fermion gas, which may further enhance the total GW signal but which also requires a more quantitative investigation in future work. Additionally, we identify a new backreaction regime in which fermion production moderates the axion--vector dynamics. In this regime, the axion velocity and all energy-density components exhibit oscillations analogous to the strong backreaction in PAI, but here, the oscillations occur around the slow-roll trajectory and are damped by the presence of charged fermions. These observations define again an interesting GEF benchmark for future lattice studies. 
}


\maketitle


\section{Introduction}
\label{sec:introduction}

The theory of cosmic inflation~\cite{starobinsky_1980, guth_1981, linde_1982, albrecht_1982, starobinsky_1982, linde_1983} has been extremely successful in explaining the large-scale homogeneity, isotropy, and spatial flatness of our present Universe
as well as the origin of the temperature anisotropies in the cosmic microwave background (CMB)~\cite{Mukhanov_1981, Mukhanov_1982,hawking_1982,guth_1982,bardeen_1983}. 
However, the simplest models of slow-roll inflation do not address how the inflaton field eventually transfers its energy to a thermal plasma of Standard Model (SM) degrees of freedom. To this end, one needs to introduce a coupling between the inflaton and the SM fields~\cite{Kofman:1997yn,Allahverdi:2010xz,Amin:2014eta}. This coupling has to be strong enough in order to allow for an efficient reheating. On the other hand, the radiative corrections to the inflaton potential due to these interactions can easily spoil its flatness and destroy the slow-roll regime of inflation. 

A model that proposes to address this issue is axion inflation~\cite{garretson_1992, anber_2006, Anber_2010}. In this model, the inflaton is an axion-like pseudoscalar coupled to gauge fields via the topological Chern--Simons 
term $\tensor{F}{_\mu_\nu} \tensor{\tilde{F}}{^\mu^\nu}$. 
This construction extends the idea of natural inflation~\cite{freese_1990} of equipping the inflaton field with an approximate shift symmetry to ensure a naturally flat potential.
In the simplest realization of axion inflation, the inflaton field couples to an Abelian vector field, but couplings to non-Abelian vector fields have also been proposed~\cite{Adshead_2012}.

The Abelian axion inflation model succeeds in sourcing a strong gauge field via a well-known tachyonic instability~\cite{garretson_1992}.
If there exist other matter fields that are charged under the gauge group in question, e.g., like in the case of the SM $U(1)_{\rm Y}$ hypercharge group, the nonperturbative phenomenon known as the Schwinger effect~\cite{Sauter_1931,Heisenberg_1936, Schwinger_1951} implies that 
the strong  (hyper)electric and (hyper)magnetic fields will create particle--antiparticle pairs of all these charged fields, i.e., every SM fermion field.
Even though the Schwinger effect has not been observed experimentally, calculations in de Sitter space~\cite{Frob_2014, Kobayashi_2014, Stahl_2015, Hayashinaka_2016_A, Hayashinaka_2016_B, Bavarsad_2016, Hayashinaka_2018, Banyeres_2018,domcke_2018,Domcke_2020_Fermions,Bastero-Gil_2025_A,Bastero-Gil_2025_B} 
indicate that this effect should also be relevant during a phase of accelerated expansion like inflation.

A refined version of axion inflation builds upon this notion by including this production mechanism for charged fields~\cite{domcke_2018, Sobol_2019, Domcke_2020_Fermions, Gorbar_2021, gorbar_2022, fujita_2022, cado_2022, domcke_2023, gorbar_2023,bastero-gil_2024_A,bastero-gil_2024_B,eckardstein_2025,iarygina_2025}.
Some references instead study the creation of charge carriers for an inflaton field coupled kinetically to the gauge field~\cite{Sobol_2018, Kitamoto_2018, Gorbar_2019, Sobol_2020}.
In both cases, charge carriers are produced in pairs and are thought to induce a conductive medium permeating the inflationary Universe, thereby damping the gauge-field production.
This effect invariably arises when the gauge bosons generated during inflation also couple to charge carriers.
Consequently, any modeling of the production of the hypercharge gauge field during inflation necessarily needs to account for this Schwinger damping.

This article is the second in a two-part series investigating the gauge field generated during Abelian axion inflation as a source for a stochastic gravitational-wave background (SGWB).
To achieve this, we employ the gradient expansion formalism (GEF)~\cite{Sobol_2019, Gorbar_2021, gorbar_2022, durrer_2023, eckardstein_2023, domcke_2024,eckardstein_2025} to determine the inflationary dynamics.
Our first article~\cite{vonEckardstein_PAI_2025} is dedicated to Abelian axion inflation without additional charge carriers coupled to the gauge field, a model we dub \textit{pure axion inflation} (PAI).
In this article, we include the effect of fermions in a model we refer to as \textit{fermionic axion inflation} (FAI). Specifically, we consider a realization of FAI in which the inflaton couples to the SM hypercharge gauge field. 
Our aim is to determine the prospects for detecting gauge-field-induced gravitational waves (GFIGW) with the Laser Interferometer Space Antenna (LISA)~\cite{LISA_2017, LISA_2019}, the Einstein Telescope (ET)~\cite{ET_2010},
or within existing pulsar timing array data sets~\cite{agazie_nanograv_2023,epta+inpta_2023,reardon_PPTA_2023,xu_CPTA_2023,miles_meerkat_2024}.
We confront these predictions with cosmological measurements, namely constraints on inflationary models from the PLANCK satellite~\cite{planck_2020_X}, and limits on the amount of additional dark relativistic degrees of freedom, $\Delta N_{\rm eff}$~\cite{planck_2020_VI, yeh_2021, pisanti_2021}.
Additionally, we account for the fact that the LIGO~\cite{LIGO_2010, LIGO_2014} and Virgo~\cite{VIRGO_2014} detector network did not observe an SGWB in their third observing run~\cite{LIGO_collaboration_2021}.

In Part I~\cite{vonEckardstein_PAI_2025}, our analysis demonstrated the difficulties with generating an observable SGWB in the PAI model.
We found that sourcing an SGWB with an appreciable amplitude such as to be observable by ET or LISA required a precise compensation between an increase in the axion--vector coupling strength and a corresponding decrease in the inflationary scale. Furthermore, we encountered a tension between an observable SGWB originating from PAI and constraints on $\Delta N_{\rm eff}$.
We identified that this conflict arises as a result of the explosive production of gauge bosons backreacting onto the inflaton dynamics, which prolongs the duration of inflation, thereby allowing for more gauge-field generation and an inevitable overproduction of gravitational radiation. As these results are based on the GEF and thus come with conceptual limitations (see below and Ref.~\cite{vonEckardstein_PAI_2025} for an extended discussion), they do not represent a general no-go theorem. Instead, our GEF benchmark study of the PAI model defines an interesting target for future numerical lattice studies. 

In the present article, we find that the tension between $\Delta N_{\rm eff}$ and an observable GFIGW signal that we had identified in the PAI model is absent for FAI coupled to the hypercharge gauge field.
Our results show that the strong (hyper)electromagnetic fields generated due to the axion coupling also produce SM fermions in great abundance, thereby draining energy from the gauge field.
This dampening of the gauge field translates to the SGWB, which is consequently only mildly blue-tilted, but is still loud enough such that it could be detected by both LISA or ET.
When these processes happen at a sufficiently high inflation scale, we even find that the signal may explain the evidence for an SGWB that the NANOGrav collaboration finds in their latest 15-year data set~\cite{agazie_nanograv_2023}.
However, this interpretation is in conflict with CMB constraints on the amplitude of scalar and tensor perturbations coming from inflation.
As a by-product of our analysis, we also show that the dampened production of gauge bosons due to fermions does not mean that the gauge field cannot backreact onto the inflaton dynamics.
When the axion--vector coupling is large enough, backreaction can still occur, however, in a much more temperate manner compared to PAI. We refer to this novel regime of backreaction as \textit{fermion-tempered backreaction}.

Evidently, fermions are crucial in FAI. They effectively suppress gauge-field production, thereby opening up a new region of parameter space which is otherwise in conflict with  $\Delta N_{\rm eff}$ constraints.
However, in this study, we model their dynamics in terms of an effective fermion current  expressed in terms of the gauge field, instead of treating them as independent degrees of freedom.
Furthermore, we show by means of basic estimates that fermions could contribute a sizeable fraction to the SGWB at high frequencies, which would affect the detection prospects for the individual GW experiments considered in this work. 
Therefore, our results motivate a more complete treatment of the contribution to the total GW signal from the fermion gas in future work.

The structure of this article is as follows: In Sec.~\ref{sec: FAI}, we review the FAI model. Sec.~\ref{sec: GWs and PT} gives a brief summary of gauge-field induced-gravitational waves.
Sec.~\ref{sec: Parameters and Constraints} discusses model constraints and GW detection characteristics.
Then, we explore fermion-tempered backreaction in Sec.~\ref{sec: Schwinger BR}, before entering an in-depth discussion of the SGWB signal sourced by FAI in Sec.~\ref{sec: Results}, where we
assume both instantaneous reheating and the impact of a lowered reheating temperature. In this last section,  we also estimate the importance of the fermions as a source for an SGWB contribution.
Sec.~\ref{sec: Conclusion} finishes with concluding remarks and an outlook. In Appendix~\ref{app: scale-dep modeling}, we compare numerical results for the SGWB spectra obtained in two different models of the Schwinger-induced damping.

\medskip\noindent
\textbf{Notation:} Throughout this paper, we assume the background spacetime to be described by the spatially flat Friedmann--Lema\^{i}tre--Robertson--Walker (FLRW) metric,
\begin{equation}
    ds^2 = \tensor{g}{_\mu_\nu} \D x^\mu \D x^\nu = \D t^2 - a^2(t) \tensor{\delta}{_i_j} \D x^i \D x^j = a^2(\eta)\left(\D \eta^2 -  \tensor{\delta}{_i_j}\D x^i \D x^j \right)\, , 
    \label{eq: FLRW}
\end{equation}
with scale factor $a$, physical time $t$, conformal time $\eta$, Greek indices for four-vectors, Latin indices for Euclidean three-vectors, and metric signature $(+,-,-,-)$.
Unless stated otherwise, $f'$ denotes the derivative of $f$ with respect to conformal time, while $\dot{f}$ denotes the derivative with respect to physical time. 
We define the Levi-Civita symbol in four dimensions such that $\tensor{\varepsilon}{^0^1^2^3} = 1$.  
All results are given in natural units, $c = \hbar = 1$, and in terms of the reduced Planck mass, $\MP = 1/\sqrt{8 \pi\,G} \simeq \SI{2.435e18}{\giga\eV}$.


\section{Fermionic axion inflation}
\label{sec: FAI}


The Lagrangian density for fermionic Abelian axion inflation describes the interactions between a $U(1)$ gauge field, $A_\mu$,
coupled via a Chern--Simons term to a pseudoscalar inflaton field, $\phi$, and interacting with fields, $\chi_i$, charged under its $U(1)$ symmetry group,
\begin{equation}
    \mathcal{L}_\mathrm{FAI} =\frac{1}{2} g^{\mu \nu} \p_{\mu} \phi \,  \p_{\nu} \phi - V(\phi) 
        - \frac{1}{4} \tensor{F}{_\mu_\nu} \tensor{F}{^\mu^\nu} - \frac{1}{4} I(\phi) \tensor{F}{_\mu_\nu} \tensor{\tilde{F}}{^\mu^\nu} 
        + \mathcal{L}_{{\rm \chi}}(\{\chi_i\}, A_{\mu}) \, .
    \label{eq: Lagrangian - FAI}
\end{equation}
In this equation, $V(\phi)$ represents the inflaton potential, $I(\phi)$ is a generic axial coupling function, and $\tensor{F}{_\mu_\nu} = \p_\mu A_\nu - \p_\nu A_\mu$ is the field-strength tensor of the Abelian gauge field.
The dual of the field strength tensor is $\tensor{\tilde{F}}{^\mu^\nu} = \tensor{\varepsilon}{^\mu^\nu^\alpha^\beta}\,\tensor{F}{_\alpha_\beta}/(2\sqrt{-g})$, with $g=\operatorname{det}g_{\mu\nu}$ being the determinant of the spacetime metric. 
The last term, $\mathcal{L}_{\chi}$, is a gauge-invariant Lagrangian density describing the interactions of the gauge field with the charge carriers $\chi_i$.%
\footnote{These charge carriers can be of arbitrary spin, including scalars, fermions, or vector bosons. However, in the particular case of the SM hypercharge gauge group $U(1)_Y$, which is of primary interest for us in this paper, fermions are the most relevant charge carriers. More specifically, we assume that the SM Higgs field is stabilized at its origin during axion inflation, e.g., via a nonminimal coupling to the Ricci scalar. In this case, electroweak symmetry is unbroken, all SM fermions are massless, but the SM Higgs field itself is heavy and hence less susceptible to Schwinger pair production. We thus restrict ourselves to fermionic charge carriers in our analysis and refer to our model as \textit{fermionic axion inflation} for precisely this reason.\label{footnote:ricci}}
As outlined below, we treat the fermions effectively; thus, we do not further specify $\mathcal{L}_{\chi}$.

By varying the action $S_{\rm FAI} = \int \D^4 x \sqrt{-g} \,\mathcal{L}_\mathrm{FAI}$ with respect to the metric, 
one derives the energy--momentum tensor for this system, which decomposes into three contributions corresponding to 
the inflaton field, the gauge field, and the charge carriers:
\begin{equation}
    \tensor{T}{_\mu_\nu} = \p_{\mu} \phi \, \p_{\nu} \phi +\tensor{F}{_\mu^\alpha} \tensor{F}{_\alpha_\nu} - g_{\mu \nu} \left(\frac{1}{2} \p_{\alpha} \phi \,  \p^{\alpha} \phi - V(\phi) - \frac{1}{4} \tensor{F}{_\alpha_\beta} \tensor{F}{^\alpha^\beta}\right) + \tensor{(T_\chi)}{_
        \mu_\nu}\, ,
    \label{eq: E--M FAI}
\end{equation}
where $\tensor{(T_\chi)}{_\mu_\nu}$ is the contribution to the energy--momentum tensor from the charge carriers.
Assuming the background spacetime to be homogeneous and isotropic, $\tensor*{T}{^\mu_\nu}$ can be separated into a background contribution, $\tensor*{\bar{T}}{^\mu_\nu}$, and a perturbation, $\tensor*{\delta{T}}{^\mu_\nu}$,
\begin{equation}
    \tensor*{T}{^\mu_\nu} = \tensor*{\bar{T}}{^\mu_\nu} + \tensor*{\delta{T}}{^\mu_\nu}, \quad 
    \tensor*{\bar{T}}{^\mu_\nu} \equiv \langle \tensor*{T}{^\mu_\nu} \rangle, \quad
    \tensor*{\delta{T}}{^\mu_\nu} \equiv \tensor*{T}{^\mu_\nu} - \langle \tensor*{T}{^\mu_\nu} \rangle\, .
    \label{eq: E--M splitting}
\end{equation}
The average $\langle \cdot \rangle$ depends on the system in question; in our case, it means taking the quantum expectation value with respect to the Bunch--Davies vacuum.
The background energy--momentum tensor can be viewed as describing an ideal fluid, whose subcomponents constitute a homogeneous energy density, $\rho$, and pressure, $p$,
\begin{equation}
       \rho = \frac{1}{2} {\dot\varphi}^2 + V(\varphi) + \frac{1}{2} \langle \bm{E}^2 + \bm{B}^2 \rangle 
            + \rho_\chi, \qquad p = \frac{1}{2} {\dot\varphi}^2 - V(\varphi) + \frac{1}{6} \langle \bm{E}^2 + \bm{B}^2 \rangle + \frac{1}{3}\rho_\chi \, .
\end{equation}
To arrive at these expressions, we assumed that spatial gradients in the inflaton field are negligible
such that only the inflaton zero mode contributes to the background expansion, $\varphi(t) = \langle \phi(t, \bm{x})\rangle$. 
Furthermore, we suppose that all charge carriers are massless, thus $p_\chi = \rho_\chi/3$.
For convenience, we also defined electric and magnetic fields, $\bm{E}$ and $\bm{B}$, via
\begin{equation}
    \tensor{F}{_0_i} = a E_i, \quad \tensor{F}{_i_j} = - a^2 \tensor{\varepsilon}{_i_j_k} B^k \, .
    \label{eq: E&B definition}
\end{equation}
These background quantities evolve according to the following coupled equations of motion,
\begin{subequations}
    \begin{equation}
        \Ddot{\varphi} + 3H\dot{\varphi} + V_{,\phi}(\varphi) = \frac{1}{2} I_{,\phi}(\varphi) \braket{\bm{E} \cdot \bm{B} + \bm{B} \cdot \bm{E}}\, ,
        \label{eq: phiEoM}
    \end{equation}
    \begin{equation}
        \diver \bm{E} = 0\, , \qquad \diver \bm{B} = 0\, ,
    \end{equation}
    \begin{equation}
        \dot{\bm{E}} + 2 H \bm{E} - \frac{1}{a}\rot \bm{B} + I_{,\phi}(\varphi)  \dot{\varphi}\bm{B} + \bm{J} = 0\, ,
        \label{eq: EEoM}
    \end{equation}
    \begin{equation}
        \dot{\bm{B}}  + 2 H \bm{B} + \frac{1}{a}\rot \bm{E} = 0\, ,
        \label{eq: BEoM}
    \end{equation} 
    \begin{equation}
        H^2 = \frac{\rho}{3 \MP^2}\,,
        \label{eq: Friedmann}
    \end{equation}
    \begin{equation}
        \dot{\rho}_\chi  + 4H\rho_\chi = \langle \bm{J} \cdot \bm{E} \rangle \,.
        \label{eq: rhoChi EoM}
    \end{equation}%
    \label{eq: EoMs}%
\end{subequations}%
The evolution equation for the energy density of the charge carriers, $\rho_\chi$, follows from covariant energy conservation at the background level.
Furthermore, we define the four-current $J^\mu = - \p \mathcal{L}_{\chi}/\p A_\mu = (\rho_c, \bm{J}/a)$, 
with $\bm{J}$ the induced current and net charge $\rho_c$. We set $\rho_c = 0$, since the Schwinger effect produces particles and antiparticles in equal shares. 

The effective description in terms of a charged current $\bm{J}$ requires further attention.
Assuming constant and (anti-)collinear classical electric and magnetic fields in an idealized de Sitter background, 
the effective current induced via Schwinger pair creation for a single particle species of charge $e|Q|$ and mass $m$ is derived to be~\cite{domcke_2018,Domcke_2020_Fermions}
\begin{equation}
    |\bm{J}_{\rm SE}'| = \frac{\left( e |Q| \right)^3}{6  \pi^2 H} |\bm{E}_{\rm c}'| |\bm{B}_{\rm c}'| \psi\left( \frac{\pi |\bm{B}_{\rm c}'|}{|\bm{E}_{\rm c}'|}\right) 
    \exp{\left(- \frac{\pi m^2}{e |Q| |\bm{E}_{\rm c}'|}\right)} \, .
    \label{eq: Schwinger current}
\end{equation}
By the subscript 'c', we indicate that the electric and magnetic fields above are classical and constant and are a priori different from $\bm{E}$ and $\bm{B}$ appearing in Eq.~\eqref{eq: EoMs}.
Quantities marked with a prime ($'$) are given in a frame where $\bm{E}_{\rm c}$ and $\bm{B}_{\rm c}$ are (anti-)collinear.
The function $\psi(x)$ is given by $\coth{x}$ or $1 / (2 \sinh{x})$ for Dirac fermions or complex scalars, respectively.
To use this non-trivial result to effectively model $\bm{J}$, we follow considerations outlined in our previous work~\cite{eckardstein_2025}.

Consider that, during axion inflation, quantum gauge-field modes are enhanced, and classicalize once their wavelengths are of the same order as the Hubble horizon.
Then, we may suppose that classical electric and magnetic fields with strengths $\langle \bm{E}^2 \rangle^{1/2}$ and $\langle \bm{B}^2 \rangle^{1/2}$, respectively, permeate a single Hubble patch
at every moment in time. Because these classical fields are also the reason for the Schwinger pair creation of charged particles, their presence also implies the existence of an effective conductive medium.
Mathematically, the effect of this medium can be quantified in terms of a conductivity $\sigma$ that relates the induced current either to $\langle \bm{E}^2 \rangle^{1/2}$ or to $\langle \bm{B}^2 \rangle^{1/2}$ via a generalized form of Ohm's law, i.e., 
\begin{equation}
    \sigma_{E}' = \frac{|\bm{J}_{\rm SE}'|}{\big\langle {\bm{E}'}^2 \big\rangle^{1/2} }\, \quad \mathrm{or} \quad \sigma_{B}' = \frac{|\bm{J}_{\rm SE}'|}{\big\langle {\bm{B}'}^2 \big\rangle^{1/2} } \,.
    \label{eq: Conductivities in the collinear frame}
\end{equation}
However, the collinearity of the electric and magnetic fields renders the precise definition of $\sigma$ a priori ambiguous. 
In the above equation, $\bm{J}'_{\rm SE}$ should be understood as computed from Eq.~\eqref{eq: Schwinger current} in terms of $\langle {\bm{E}'}^2 \rangle$ and $\langle {\bm{B}'}^2 \rangle$.
Since the quantum electric and magnetic fields also feel this conductivity, 
these considerations motivate a first definition of the quantum induced current $\bm{J}$ entering in our EOMs:
\begin{equation}
    \bm{J} \equiv \sigma_{E} \bm{E} + \sigma_{B} \bm{B}\, .
    \label{eq: Ohmic current}
\end{equation}
That we consider both magnetic and electric conductivities in this definition is a consequence of the fact that the electric and magnetic fields are not necessarily collinear in the comoving frame in which Eq.~\eqref{eq: Ohmic current} applies.
The conductivities $\sigma_{E/B}$ may be derived by boosting to the collinear frame, computing the conductivities from Eq.~\eqref{eq: Schwinger current}, 
and boosting back to the comoving frame. This procedure yields
\begin{align}
    \sigma_{E} &= \sqrt{\frac{\langle \bm{E}^2 \rangle - \langle \bm{B}^2 \rangle + \Sigma }{ \langle \bm{E}^2 \rangle + \langle \bm{B}^2 \rangle + \Sigma } }
    \sum_{i} \mathfrak{J}_i\, , \\
    \sigma_{B} &= \operatorname{sign}{ \left( \langle \bm{E} \cdot \bm{B}\rangle \right) }
    \sqrt{\frac{\langle \bm{B}^2 \rangle - \langle \bm{E}^2 \rangle + \Sigma }{ \langle \bm{E}^2 \rangle + \langle \bm{B}^2 \rangle + \Sigma }}\sum_{i} \mathfrak{J}_i \, ,
    \label{eq: Conductivities non-collinear}
\end{align}
where $\Sigma = \sqrt{ [\langle\bm{E}^2\rangle - \langle\bm{B}^2\rangle]^2 + 4 \langle \bm{E} \cdot \bm{B} \rangle^2 }$. The $\mathfrak{J}_i$ encompass the contributions to $\sigma_{E/B}$ of the individual charged fermions and scalars, $\chi_i$, with respective masses $m_i$ and charges $Q_i$
\begin{equation}
    \mathfrak{J}_i = \frac{ \left(e |Q_i|\right)^3 }{6  \pi^2 H} \frac{|\langle \bm{E}\cdot\bm{B} \rangle|}{\sqrt{\Sigma}}
    \psi_i\left( \pi \sqrt{\frac{\langle \bm{B}^2 \rangle - \langle \bm{E}^2 \rangle + \Sigma }{ \langle \bm{E}^2 \rangle - \langle \bm{B}^2 \rangle + \Sigma }}\right)  
    \exp{ \left(- \frac{\pi m^2_i \sqrt{2}}{e |Q_i| (\langle \bm{E}^2 \rangle- \langle \bm{B}^2 \rangle + \Sigma)^{1/2}}\right) } \, .
    \label{eq: current fraction}
\end{equation}

The Ohmic form of the Schwinger-induced current in Eq.~\eqref{eq: Ohmic current} implies that all gauge-field modes are equally affected by the conductive medium.
However, as we argued in Ref.~\cite{eckardstein_2025}, this description cannot reflect the realistic physical situation. In this simple picture, gauge-field modes with wavelengths much shorter than the typical
separation of charge carriers in the conductive medium would feel its effect just as much as modes with far greater wavelengths.
To mitigate the severity of this assumption, we account for an additional scale dependence in Eq.~\eqref{eq: Ohmic current} by making the following modification,
\begin{equation}
    \bm{J}(t, \bm{x}) \equiv \frac{1}{(2 \pi)^{3}}\int \D^3 \bm{y}\, \Big(\sigma_{E}(t) \bm{E}(t, \bm{y}) + \sigma_{B}(t) \bm{B}(t, \bm{y})\Big) \int \D^3 \bm{k} \, \Theta(t, k) e^{i \bm{k}\cdot (\bm{x} - \bm{y})} \, .
    \label{eq: ScaleDep Ohmic current}
\end{equation}
By convoluting the original current $\bm{J}$ with a heuristic function $ \int \D^3 \bm{k} \, \Theta(t,k)e^{i \bm{k}\cdot \bm{x}}$,
we gain the flexibility to reflect the spectral dependence of the conductive medium by appropriately choosing $\Theta(t, k)$.
In Ref.~\cite{eckardstein_2025}, we discussed a well-motivated choice for this spectral dependence. This relies on identifying a characteristic momentum scale, $k_{\rm S}(t)$,
which reflects the typical particle--antiparticle separation following their creation. This scale may be estimated from the strength of the electric field at time $t$, $k_{\rm S}/a \propto |{\bm{E}'_{\rm c}}|^{1/2}$.
We give more details on this choice for  $\Theta(t, k)$ in Appendix~\ref{app: scale-dep modeling}.

It is instructive to study the impact of the conductive medium in Fourier space. To this end, consider the gauge field $A_\mu$ in radiation gauge ($\diver \bm{A} = 0$, $A_0=0$),
\begin{equation}
    \hat{\bm{A}}(\eta, \bm{x}) = \int \frac{\D^3 \bm{k}}{(2 \pi)^{3/2}}\sum_{\lambda={\pm1}} 
    \left( \bm{\epsilon}_\lambda(\bm{k}) A_\lambda(\eta, k) \hat{a}^{\vphantom{\dagger}}_\lambda(\bm{k})e^{i \bm{k} \cdot \bm{x}} 
    + \bm{\epsilon}_\lambda^*(\bm{k}) A^*_\lambda(\eta, k) \hat{a}_\lambda^\dagger(\bm{k})e^{-i \bm{k} \cdot \bm{x}} \right)\, ,
    \label{eq: GF FT}
\end{equation}
in a helicity basis with polarization vectors $\bm{\epsilon}_\lambda(\bm{k})$ obeying
\begin{align}
\bm{k} \cdot \bm{\epsilon}_\lambda(\bm{k}) &= 0, &\,i \bm{k} \times \bm{\epsilon}_\lambda(\bm{k}) &= \lambda k\, \bm{\epsilon}_\lambda(\bm{k}), \nonumber\\
    \bm{\epsilon}_\lambda(\bm{k})\cdot\bm{\epsilon}_{\lambda'}^*(\bm{k}) &= \delta_{\lambda \lambda'}, &\,\bm{\epsilon}^\ast_\lambda(\bm{k})=\bm{\epsilon}_{-\lambda}(\bm{k})&=\bm{\epsilon}_{\lambda}(-\bm{k}) \, ,
    \label{eq: polarisation vectors}
\end{align}
and creation and annihilation operators $\hat{a}^\dagger_\lambda(\bm{k})$ and $\hat{a}^{\vphantom{\dagger}}_\lambda(\bm{k})$ satisfying
\begin{equation}
    [\hat{a}^{\vphantom{\dagger}}_\lambda(\bm{k}), \hat{a}^\dagger_{\lambda'}(\bm{k'})] = \delta_{\lambda \lambda'} \delta^{(3)}(\bm{k} - \bm{k'}), \quad
     [\hat{a}^{\vphantom{\dagger}}_\lambda(\bm{k}), \hat{a}^{\vphantom{\dagger}}_{\lambda'}(\bm{k'})] = [\hat{a}^\dagger_\lambda(\bm{k}), \hat{a}^\dagger_{\lambda'}(\bm{k'})] = 0 \, .
    \label{eq: canonical commutation relation}
\end{equation}
From Amp\`{e}re's law in Eq.~\eqref{eq: EEoM} and the induced current $\bm{J}$ in Eq.~\eqref{eq: ScaleDep Ohmic current},
one may derive an evolution equation for the mode functions $A_\lambda(t, k)$,
\begin{equation}
    \ddot{A}_\lambda(t, k)  + \big( H + \sigma_{E} \Theta(t, k) \big) \dot{A}_\lambda(t, k) 
    + \left[ \left(\frac{k}{a}\right)^2 - \lambda \frac{k}{a} \big( 2 \xi H + \sigma_{B}\Theta(t, k) \big) \right] A_\lambda(t, k) = 0 \,,
    \label{eq: Mode Eq - FAI}
\end{equation}
which reduces to the evolution equation in the PAI model for $\sigma_{E/B} = 0$. Here, as usual, $\xi=I_{,\phi}(\varphi)\dot{\varphi}/(2H)$ denotes the instability parameter or gauge-field production parameter of axion inflation, which, in the absence of fermions, determines the threshold momentum below which gauge-field modes experience the tachyonic enhancement.
Qualitatively, the physical interpretation of this equation matches the one of PAI:
gauge-field production is a consequence of the non-zero inflaton velocity, $\dot{\varphi}$,
which triggers a tachyonic instability in Eq.~\eqref{eq: Mode Eq - FAI} for one of the gauge-field helicities, $\lambda=\pm1$,
depending on the sign of $\dot{\varphi}$. 
However, the Schwinger effect reduces the efficiency of gauge-field production, as $\sigma_{B}$ takes an opposite sign to $\xi$, 
yielding an impeded effective instability parameter $\xi_{\rm eff} = \xi + \sigma_{B}/(2H)$. Meanwhile, the electric conductivity $\sigma_{E}$ dampens the gauge-field modes by adding to Hubble friction.
The role of the heuristic spectral function $\Theta(t,k)$ is also apparent in Eq.~\eqref{eq: Mode Eq - FAI}: it determines the range of gauge-field momenta $k$ that are subject to the influence of the purely time-dependent damping terms $\sigma_{E/B}$.

It is important to note the ramifications for the dynamics and phenomenology of axion inflation because of the inhibited gauge-field generation due to the Schwinger effect.
The main difference regarding the dynamical evolution is a decrease in $\langle \bm{E} \cdot \bm{B}\rangle$ such that the backreaction onto the inflaton
via the friction term on the right-hand side in Eq.~\eqref{eq: phiEoM} is diminished. 
For a particular benchmark point, Ref.~\cite{gorbar_2023} demonstrated that a PAI system experiences a phase of strong gauge-field backreaction while a comparable FAI system did not.
To understand the implication of this, consider that strong backreaction results in an increased duration of inflation by drawing kinetic energy from the inflaton field and instead allowing for more gauge-field production, 
which ultimately ends inflation via $V(\varphi) \sim \rho_{EM}$ instead of $V(\varphi) \sim \dot{\varphi}^2$. 
Therefore, decreasing the impact of backreaction implies that, in FAI, inflation ends on the slow-roll trajectory with a subdominant energy density in $\rho_{EM}$.
This implies that an extended phase of reheating may still take place after the end of axion inflation.
A second implication of absent backreaction regards axion inhomogeneities, which we have neglected throughout the derivation presented above.
In PAI, these inhomogeneities become large during a phase of strong gauge-field backreaction and can alter the dynamical evolution of the inflaton--gauge-field system~\cite{Figueroa_2023, Figueroa_2024,caravano_2023,sharma_2025}. 
At present, this means that studies of PAI that neglect these inhomogeneities, like the GEF, can only qualitatively capture the phase of strong backreaction~\cite{Figueroa_2023,Figueroa_2024}.
To obtain the full dynamics of PAI, lattice simulations are necessary.
However, as we expect an overall reduced dynamical impact of the gauge field during FAI, it is conceivable that also fewer axion inhomogeneities are sourced.

{In this regard, we would like to mention the recent work~\cite{iarygina_2025}, which studies the same physical model by solving the corresponding equations of motion on a lattice using the \textsc{Pencil Code}~\cite{PencilCode:2020eyn}. The impact of the Schwinger effect is incorporated by employing the same Ohmic form of the Schwinger current~\eqref{eq: Ohmic current} with the conductivities given in~\eqref{eq: Conductivities non-collinear}. One important difference between their lattice implementation and the approach used in the present work concerns the treatment of Schwinger damping of the gauge-field modes. Here, we account for the damping only of those modes whose wavelengths exceed the Schwinger pair-production length scale. In contrast, Ref.~\cite{iarygina_2025} applies the same damping to all gauge-field modes, independently of their momentum, since it is not feasible to perform efficient mode selection in position space on a lattice.
With this assumption, the authors of Ref.~\cite{iarygina_2025} carried out a numerical analysis at three benchmark points in parameter space (specifically, for the axion mass $m=5.2\,\times\,10^{-6}\,M_{\rm P}$ and axion--vector coupling $\beta=12,\,15,$ and $18$).}%
\footnote{{Note that a direct comparison of our results and the ones of Ref.~\cite{iarygina_2025} is impossible since we are considering higher values of the coupling constants, $\beta\geq 20$ which are needed to obtain a sufficiently strong GW signal.}}
{They found excellent agreement between the fully inhomogeneous simulations and the homogeneous setup where axion gradients were manually switched off. This outcome confirms our expectations discussed above, at least within the parameter region explored in Ref.~\cite{iarygina_2025}. Qualitatively, we see no reason for axion gradients to become important at larger coupling values unless the backreaction eventually becomes significant.
In Sec.~\ref{sec: Schwinger BR}, we identify and investigate the properties of this new regime of \textit{fermion-tempered backreaction}, still within the homogeneous approximation. These results provide a GEF benchmark for this region of parameter space, and they should be validated by future lattice simulations.}

{Generally speaking, lattice simulations have two main advantages compared to semianalytical methods:
(i)~they can naturally incorporate axion inhomogeneities, and
(ii)~they allow one to study gauge-field and fermion production after the end of inflation, during the preheating stage.
On the other hand, semianalytical approaches such as the GEF offer several important advantages:
(a)~they are not limited in dynamical range or simulation duration;
(b)~they allow one to treat different gauge-field modes differently (for example, to include selective damping or to isolate partial contributions to the energy densities);
(c)~they are significantly faster and enable parameter scans over wide regions of the model space, rather than exploring only isolated benchmark points.
Moreover, lattice methods often rely on the output of semianalytical homogeneous approaches for calibration purposes and for obtaining accurate initial conditions for the simulation. Therefore, we believe that combining our GEF results with future lattice simulations will lead to a highly synergistic outcome.}

%

Qualitatively, the obstructed gauge-field production also inhibits the subsequent sourcing of GFIGWs.
On the other hand, the energy from the gauge field is instead transferred to the charge carriers, as may be seen from Eq.~\eqref{eq: rhoChi EoM}.
Previous studies found this production of charge carriers to be efficient enough such that $\rho_{\chi}$ surpasses $\rho_{\rm EM}$ at the end of inflation~\cite{Gorbar_2021, vonEckardstein_PAI_2025}.
If the fluid associated with these charge carriers is anisotropic, it will add a second induced component to the GW spectrum.
%


\section{Tensor power spectrum and gravitational waves}
\label{sec: GWs and PT}


We study the production of tensor perturbations by working with a first-order perturbed FLRW metric,
tracking only the transverse and traceless tensor perturbations, $\tensor*{h}{_i_j^{\mathrm{TT}}}$,
\begin{equation}
    ds^2 = a^2(\eta) \left( \D \eta^2-  \left[\tensor{\delta}{_i_j} + \tensor*{h}{_i_j^{\mathrm{TT}}}\right]\D x^i \D x^j \right) \, .
\end{equation}
We describe these perturbations as a quantum field, which we can express in Fourier space using a circular polarisation basis such that
\begin{equation}
    \tensor*{\hat{h}}{_i_j^{\mathrm{TT}}}(\eta, \bm{x}) 
    = \int \frac{\D^3 \bm{k}}{(2 \pi)^{3/2}} \sum_{\lambda={\pm1}}\left( \epsilon_i^\lambda(\bm{k}) \epsilon_j^\lambda(\bm{k}) \hat{h}_\lambda(\eta, k) e^{i \bm{k} \cdot \bm{x}} + \text{h.c.} \right)\, .
    \label{eq:hijTT}
\end{equation}
where the polarization vectors $\bm{\epsilon}^\lambda(\bm{k})$ are defined as in Eq.~\eqref{eq: polarisation vectors}. The Fourier amplitudes, $\hat{h}_\lambda(\eta, k)$, evolve according to
\begin{equation}
    \hat{h}_\lambda'' + 2 \mathcal{H} \hat{h}_\lambda' + k^2 \hat{h}_\lambda =
     \frac{2}{\MP^2} a^2 \tensor*{\Pi}{_\lambda^{ij}}(\bm{k})  \int \frac{\D^3 \bm{x}}{(2 \pi)^{3/2}} \tensor{\sigma}{_i_j}  e^{-i \bm{k}\cdot\bm{x}} \, ,
    \label{eq: GW EoM, Fourier}
\end{equation}
where $\mathcal{H} = a'/a$ is the comoving Hubble rate, $\tensor*{\Pi}{_\lambda^{ij}} = \epsilon^i_{-\lambda}(\bm{k}) \epsilon^j_{-\lambda}(\bm{k})$ is the projector onto the helicity basis
and $\tensor{\sigma}{_i_j}$ is the anisotropic stress of the perturbed energy--momentum tensor $\delta \tensor*{T}{^\mu_\nu}$.
In the case of FAI, we derive the anisotropic stress from Eq.~\eqref{eq: E--M FAI}, finding
\begin{equation}
    \tensor*{\sigma}{_i_j} = \p_i \phi \p_j \phi - \left(E_i E_j + B_i B_j \right) +  \tensor*{\sigma}{_i_j^{\mathrm{\chi}}}\, .
    \label{eq: anisotropic stress - FAI}
\end{equation}
Assuming axion inhomogeneities to be small, we may neglect the first term. The second term, the anisotropic stress due to the gauge field, is sourced by the tachyonic instability in Eq.~\eqref{eq: Mode Eq - FAI}, 
and is analogous to the one obtained in PAI. The final term, $\tensor*{\sigma}{_i_j^{\mathrm{\chi}}}$, reflects the anisotropic stress of the charged fields $\chi_i$. 
As before, we keep this term generic, as we are treating all charged fields effectively without resolving their dynamics.

From here, we proceed as in our preceding article, Ref.~\cite{vonEckardstein_PAI_2025}, writing the solution to Eq.~\eqref{eq: GW EoM, Fourier} in terms of a homogeneous (vacuum) and a particular (induced) solution,
$\hat{h}_\lambda = \frac{2}{\MP}\left(\hat{u}_\lambda^{\mathrm{vac}} +  \hat{u}_\lambda^{\mathrm{ind}}\right)$.
In the case of FAI, the induced contribution further separates into two parts, $\hat{u}_\lambda^{\mathrm{ind}} = \hat{u}_\lambda^{\mathrm{GF}} + \hat{u}_\lambda^{\mathrm{\chi}}$, one induced by the gauge field, the other by the charged fields.
Defining the tensor power spectrum for the polarization $\lambda = \pm 1$ as
\begin{equation}
    \langle \hat{h}_\lambda(\eta,\bm{k}) \hat{h}_{\lambda'}(\eta,\bm{k'}) \rangle 
    = \delta^{(3)}(\bm{k} + \bm{k'}) \delta_{\lambda\lambda'} \frac{\pi^2}{k^3} \mathcal{P}_{T,\lambda}(k) \,,
    \label{eq: polarised PT}
\end{equation}
the total tensor power spectrum, $\mathcal{P}_{T}(k) = \sum_ {\lambda \pm1} \mathcal{P}_{T,\lambda}(k)$, receives three contributions,
\begin{equation}
    \mathcal{P}_{T}(k) = \mathcal{P}_{T}^{\rm vac}(k) + \mathcal{P}_{T}^{\rm GF}(k) + \mathcal{P}_{T}^{\rm \chi}(k)\,,
    \label{eq: PT total}
\end{equation}
assuming that there are no cross-terms between $\hat{u}_\lambda^{\mathrm{GF}}$ and $\hat{u}_\lambda^{\mathrm{\chi}}$. 
This separation is expected from Wick's theorem, as both the charged fields and the gauge field are expressed in terms of their own set of annihilation and creation operators.
We presented a more detailed derivation of the first two contributions in Part I, Ref.~\cite{vonEckardstein_PAI_2025}, the result of which is
\begin{subequations}
    \begin{align}
         \mathcal{P}_{T}^{\mathrm{vac}}(\eta, k) &= \frac{4 k^3}{\pi^2 \MP^2} |u_\lambda^0(\eta, k)|^2\, ,\\
         \mathcal{P}_{T}^{\mathrm{GF}}(\eta, k) &= \sum_{\lambda = \pm 1}\frac{k^3}{2 \pi^2 \MP^4} \int \frac{\D^3 \bm{p}}{(2 \pi)^3} \sum_{\alpha,\beta = \pm1} 
            \left(1 +  \lambda \alpha \frac{\bm{k} \cdot \bm{p}}{k p} \right)^2 \left(1 +  \lambda \beta \frac{k^2 - \bm{k} \cdot \bm{p}}{kq}  \right)^2 \\ 
        &\times \left|\int_{-\infty}^0 \D \tau \frac{G_k(\eta, \tau)}{a^2(\tau)} 
            \left[A'_\alpha(\tau, p)A'_\beta(\tau, q) + \alpha \beta\, p q\, A_\alpha(\tau, p) A_\beta(\tau, q) \right] \right|^2 \nonumber \, .
    \end{align}%
    \label{eq: PT induced}%
\end{subequations}%
Here, the gauge-field modes $A_\lambda(t,k)$ are solutions to Eq.~\eqref{eq: Mode Eq - FAI}, $G_k(\eta, \tau)$ is the retarded Green function to the differential operator $\mathcal{D} =  (\p^2/\p \eta^2) + 2 \mathcal{H} (\p/\p \eta) + k^2$,
and $q = |\bm{k} - \bm{p}|$.
A detailed computation of the $\chi$-induced contribution, $\mathcal{P}_{T}^{\rm \chi}(k)$, is beyond the scope of the current paper. 
Therefore, we will simply keep this term generic, opting for a simple scaling argument to estimate its importance in the later part of this article.

From the tensor power spectrum, one can directly compute the GW energy density per logarithmic frequency interval in units of the critical energy density,
\begin{equation}
    \Omega_\mathrm{GW}(f) \equiv  \frac{1}{3 H_0^2 \MP^2}\frac{\D \rho_\mathrm{GW} (f)}{\D \ln{f}} = \frac{\pi^2}{3 H_0^2} f^2 |\mathcal{T}_{\rm GW}(f)|^2 \mathcal{P}_T(\eta_{\mathrm{out}}(k_f), k_f) \, , \quad k_f = 2 \pi a_0 f \, ,
    \label{eq: omega-GW from PT}
\end{equation}
where $H_0 = h \times \SI{100}{\kilo\metre\per\second\per\mega\parsec}$ is the Hubble constant and the transfer function $\mathcal{T}_{\rm GW}$ describes the evolution of a GW mode with momentum $k_f$ from horizon re-entry when $k_f \simeq \mathcal{H}(\eta_{\mathrm{in}})$ (after horizon exit during inflation when $k_f \simeq \mathcal{H}(\eta_{\mathrm{out}})$) until today~\cite{caprini_2018},
\begin{equation}
    |\mathcal{T}_\mathrm{GW}(f)|^2 \simeq \frac{H_0^2\Omega_r}{8 \pi^2 f^2} \frac{g_*(T_f)}{g_*(T_0)} \left(\frac{g_{*,S}(T_0)}{g_{*,S}(T_f)}\right)^{4/3} \left[1 + \frac{9}{16}\left(\frac{f_{\rm eq}}{\sqrt{2} f} \right)^2\right] |\mathcal{T}_{\rm reh}(f)|^2 \,.
    \label{eq:transfer}
\end{equation}
The functions $g_*(T)$ and  $g_{*,S}(T)$ represent the effective number of relativistic degrees of freedom in the SM thermal plasma at temperature $T$ contributing to the energy density and entropy, respectively.
The temperature $T_f$ corresponds to the time when $k_f$ re-entered the horizon and $T_0 \simeq \SI{2.73}{K}$ is the temperature of the CMB photons today.
$\Omega_r$ is the fractional energy density in radiation today, $h^2\Omega_r = \num{4.2e-5}$.
The fourth factor describes the transition from radiation to matter domination, with $f_{\rm eq} = \SI{2.1e-17}{\hertz}$ the frequency corresponding to matter--radiation equality.
The last factor, $|\mathcal{T}_{\rm reh}(f)|^2$, affects modes which re-enter the horizon during a period of reheating before the onset of radiation domination.
In our previous article, Ref.~\cite{vonEckardstein_PAI_2025}, we assumed instantaneous reheating, thus $|\mathcal{T}_{\rm reh}(f)|^2 = 1$. 
However, as discussed in the previous section, FAI leaves more room for a period of reheating, which can impact the GW spectrum via this transfer function.

Reheating implies that frequencies experience an additional redshift due to this non-standard phase of cosmological expansion.
This is evident when expanding the relationship $k_f = 2 \pi a_0 f$, between a frequency $f$ and the corresponding comoving momentum $k_f$,%
\footnote{We fix the normalization of the scale factor during the early stages of inflation. In this convention, the present-day value of the scale factor $a_0$ is not fixed to a certain value (e.g., $a_0 =1$), but depends on the details of the expansion history. For our purposes, this convention proves advantageous, since it allows us to use the wavenumber $k_f$ as a clock variable during inflation that is independent of the details of reheating.}
\begin{equation}
    f = \frac{k_f}{2 \pi a_0} = \frac{k_f}{2 \pi a_{\rm end}} \frac{a_{\rm end}}{a_{\rm reh}} \frac{a_{\rm reh}}{a_0} = \frac{k_f}{2 \pi a_{\rm end}} e^{-N_{\rm reh}} \left( \frac{g_{*,S}(T_0)}{g_{*,S}(T_{\mathrm{reh}})}\right)^{1/3} \frac{T_0}{T_{\mathrm{reh}}} \,,
    \label{eq: f from k}
\end{equation}
where $a_{\rm end}$ denotes the end of inflation, which is followed by $N_{\rm reh} = \ln (a_{\rm reh}/a_{\rm end})$ $e$-folds of reheating
until radiation domination sets in at a temperature $T_{\rm reh}$.
By modeling reheating as an expansion phase governed by a non-interacting fluid with equation of state $w_{\rm reh}$, the duration of reheating, $N_{\rm reh}$, may be related to the reheating temperature, $T_{\rm reh}$, via
\begin{equation}
    N_{\rm reh} = \frac{1}{3(1 + w_{\rm reh})} \ln \left(\frac{90 \MP H_{\rm end}^2}{\pi^2 g_*(T_{\rm reh}) T_{\rm reh}^4} \right) \,.
\end{equation}
Assuming a standard reheating mechanism where the dominant energy density is in inflaton oscillations around its potential minimum, reheating will manifest itself as a period of early matter domination, $w_{\rm reh} = 0$.
For this case, the transfer function $|\mathcal{T}_{\rm reh}(f)|^2$ has been computed in Refs.~\cite{Kuroyanagi_2014,Kuroyanagi_2020},
\begin{equation}
    |\mathcal{T}_{\rm reh}(f)|^2 = \frac{ \theta(f_{\rm end} - f)}{1 - 0.22 \left(\frac{f}{f_{\rm reh}}\right)^{1.5} + 0.65 \left(\frac{f}{f_{\rm reh}}\right)^{2}} \,,
    \label{eq: Transfer RH}
\end{equation}
with the frequencies $f_{\rm end}$ and $f_{\rm reh}$ corresponding to the end of inflation and the end of reheating, respectively.
As evident from Eqs.~\eqref{eq: f from k} and ~\eqref{eq: Transfer RH}, reheating affects the GW spectrum by further redshifting all frequencies with respect to instantaneous reheating,
while also suppressing the GW spectrum at high frequencies, $f \gtrsim f_{\rm reh}$, by a factor $f^{-2}$.


\section{Model specifications and constraints}
\label{sec: Parameters and Constraints}


For the remainder of this article, we focus on the scenario where the gauge field $A_\mu$ in Eq.~\eqref{eq: Lagrangian - FAI} is the SM hypercharge field.
The charge carriers $\chi_i$ therefore correspond to the fermions of the SM, which we assume to be massless; see the discussion in Footnote~\ref{footnote:ricci}. 
These assumptions are expressed in the individual fermion contributions $\mathfrak{J}_i$ to the conductivities $\sigma_{E/B}$ in Eq.~\eqref{eq: Conductivities non-collinear}, which sum up to
\begin{equation}
    \sum_{i} \mathfrak{J}_i = C\, \frac{ g'(\mu)^3 }{6  \pi^2 H} \frac{|\langle \bm{E}\cdot\bm{B} \rangle|}{\sqrt{\Sigma}}
    \operatorname{coth}\left( \pi \sqrt{\frac{\langle \bm{B}^2 \rangle - \langle \bm{E}^2 \rangle + \Sigma }{ \langle \bm{E}^2 \rangle - \langle \bm{B}^2 \rangle + \Sigma }}\right)  \, ,
\end{equation}
with $C=41/12$, equal to half the sum of the cubes of hypercharges of all SM fermions. The coupling $g'(\mu)$ is the running hypercharge gauge coupling at one-loop order,
\begin{equation}
    g'(\mu) =  g'(m_Z)  \left(1 + g'(m_Z)^2 \frac{41}{48 \pi^2} \ln \frac{m_Z}{\mu}  \right)^{-1/2}  \, ,
\end{equation}
with the $Z$-boson mass, $m_Z \simeq \SI{91.2}{\giga\eV}$, $g'(m_Z) \simeq 0.35$ and the characteristic energy scale of pair creation $\mu =\rho_{\rm EM}^{1/4}$.
For brevity, we will refer to this realization of FAI as \FAISM.%
\footnote{{Reference~\cite{Ferreira:2017lnd} argues that gauge-field modes can efficiently interact with each other via fermions and thus thermalize already when the instability parameter satisfies $|\xi|\gtrsim 2.9$ for Standard Model couplings. This argument is based on a computation of the rate for the Breit--Wheeler process $\gamma\gamma\to e^{+}e^{-}$, enhanced by the large occupation numbers of the gauge-field modes.
Although we agree that the computation itself may be correct, we believe that its interpretation is misleading. Gauge-field modes with large occupation numbers correspond to a superhorizon classical gauge field rather than on-shell photons, and the computed rate effectively describes the Schwinger pair-production process. Indeed, our numerical solutions confirm that Schwinger pair production becomes efficient when $|\xi_{\rm eff}|$ is greater than 3. Note also that the inverse process $e^{+}e^{-}\to \gamma\gamma$ is not efficient and does not lead to thermalization, since fermionic states cannot achieve large occupation numbers. Therefore, throughout this work, we assume that the gauge field does not thermalize.}}

Just as in the analysis of PAI in Part I~\cite{vonEckardstein_PAI_2025}, we work with a linear axion--vector coupling, $I(\phi) = (\beta/\MP)\,\phi$
and assume a simple quadratic scalar potential $V(\phi) = m^2 \phi^2 /2$. As we argued in Part I, this quadratic potential is representative of a larger class of inflationary models, whose late-time attractor is nearly quadratic below some threshold field value $\varphi_{\rm thr}$.
Inflation simulated on a quadratic potential yields representative GW spectra for frequencies $f \gtrsim f_{\rm thr}$, where $f_{\rm thr}$ is the frequency corresponding to modes which exit the horizon when $\varphi = \varphi_{\rm thr}$.
Assuming that the inflaton trajectory for $\varphi < \varphi_{\rm thr}$ is approximately given by the slow-roll attractor of chaotic inflation, one may deduce the relationship
\begin{equation}
    V(\varphi_{\rm thr}) \simeq m^2 \MP^2 \left[ 20 - 4.6 \log_{10} \left(\frac{f_{\rm thr}}{\SI{}{\hertz}}\right) 
    + 1.54 \log_{10} \left(\frac{m}{\MP} \frac{T_{\rm reh}}{\SI{}{\giga\electronvolt}}\right)
    -2 \Delta N_{\rm BR}\right] \, , 
    \label{eq: thr potential}
\end{equation}
which is the generalization of Eq.~(4.5) in Ref.~\cite{vonEckardstein_PAI_2025} that allows for a period of reheating with $w_{\rm reh} = 0$.
We use this relationship to determine bounds on $m$ by using the measured value 
of the amplitude of scalar perturbations, $A_S = \num{2.1e-9}$ \cite{planck_2020_VI,galloni_2024}, bounds on the tensor-to-scalar ratio, $r \lesssim 0.03$, \cite{tristram_2022, galloni_2024},
and non-Gaussianity at CMB scales, $|\xi_{\mathrm{CMB}}| \lesssim 2.5$~\cite{Barnaby_2011_A,Barnaby_2011_B,Planck:2015zfm,Planck:2019kim},
\begin{subequations}
    \begin{align}
    V(\varphi_{\mathrm{thr}}) &\lesssim  \frac{3}{2} \pi^2 A_S r \MP^4 \lesssim \num{9.3e-10} \MP^4\,. \\
    V(\varphi_{\mathrm{thr}}) &\lesssim  48 \pi^2 A_S \frac{|\xi_{\mathrm{CMB}}|}{\beta^2}\MP^4 \lesssim \num{6.2e-6} \MP^4/\beta^2 \,.
    \label{bounds}
    \end{align}%
    \label{eq: CMB bounds}%
\end{subequations}%
The second bound is only more constraining for $\beta \sim \numrange[range-phrase = -]{80}{100}$ (which we never reach in the present study), so we focus entirely on the first bound.
Modeling $V(\varphi_{\mathrm{thr}})$ as in Eq.~\eqref{eq: thr potential}, we can observe that the constraint on $m$ is marginally relaxed when reducing $f_{\rm thr}$ 
with variations ranging from $m \lesssim (\numrange[range-phrase = -]{3.1}{3.4})\times \num{e-6} \MP$ for $f_{\rm thr} = \numrange[range-phrase = -]{e-12}{e-9}\,\SI{}{\hertz}$ and instantaneous reheating.
A lower reheating temperature also only mildly affects the bounds: even for $T_{\rm reh} \simeq \SI{1}{\mega\electronvolt}$, we only find $m \lesssim (\numrange[range-phrase = -]{3.9}{4.4})\times\num{e-6} \MP$ for $f_{\rm thr} = \numrange[range-phrase = -]{e-12}{e-9}\,\SI{}{\hertz}$.
The largest effect can be obtained by allowing for an extremely prolonged duration of inflation due to backreaction. For example, for $\Delta N_{\rm BR} = 20$ and varying $f_{\rm thr} = \numrange[range-phrase = -]{e-12}{e-9}\,\SI{}{\hertz}$,
one obtains $m \lesssim (\numrange[range-phrase = -]{4.1}{4.8})\times\num{e-6} \MP$ assuming instantaneous reheating, and $m \lesssim \numrange[range-phrase = -]{6.4e-6}{e-5}\MP$ for $T_{\rm reh} \simeq \SI{1}{\mega\electronvolt}$.
Evidently, only when allowing for a low reheating temperature and many $e$-folds of backreaction can the bound on $m$ be substantially relaxed.
In practice, we will indicate bounds for $\Delta N_{\rm BR}=0$ in our results while still studying inflaton masses as high as $\num{5e-5} \MP$ due to the uncertainties involved in these estimates.

On the side of GWs, we follow the same prescription as in our companion paper.
The integrated GW energy density is constrained by Big Bang nucleosynthesis (BBN) and CMB measurements of the effective number of relativistic degrees of freedom beyond those of the SM.
We denote this excess above $N_{\rm eff}^{\rm SM}$~\cite{drewes_2024} as
\begin{equation}
    \Delta N_{\mathrm{eff}} = N_{\mathrm{eff}} - N_{\mathrm{eff}}^{\mathrm{SM}}  , \qquad N_{\mathrm{eff}}^{\mathrm{SM}} =  3.0440 \pm 0.0002 \, .
\end{equation}
In principle, we use a conservative limit $\Delta N_{\rm eff} \lesssim 0.5$ as suggested in Ref.~\cite{agazie_nanograv_RPL_2025}, accounting for the uncertainty of combined BBN and CMB constraints~\cite{planck_2020_X,yeh_2021,pisanti_2021}.
In practice, however, we find this limit to be irrelevant in this study, as the corresponding bound on $h^2 \Omega_\mathrm{GW}(f)$,
\begin{equation}
    \int_{f_{_\mathrm{BBN}}}^{f_{\mathrm{end}}} \frac{\D f}{f} h^2 \Omega_\mathrm{GW}(f) \lesssim \num{5.6e-6} \Delta N_{\mathrm{eff}} \, ,
    \label{eq: Neff}
\end{equation}
is never violated.

We estimate the detection prospects for a GW signal by a given GW observatory using the signal-to-nose ratio (SNR)~\cite{allen_1996, allen_1997, maggiore_2000},
\begin{equation}
    S/N = \left(n_{\mathrm{det}} t_{\mathrm{obs}} \int_{f_{\mathrm{min}}}^{f_{\mathrm{max}}} \D f \, \left( \frac{\Omega_{\rm signal}(f)}{\Omega_{\rm noise}(f)}\right)^2 \right)^{1/2}\, .
    \label{eq: SNR}
\end{equation}
The detector sets the frequency band, $[f_{\mathrm{min}}, f_{\mathrm{max}}]$, and if the SGWB search is based on an auto-correlation or a cross-correlation measurement, $n_{\mathrm{det}} = 1$ or $2$.
For detectors that are not yet operational, we set the observation time to $t_{\mathrm{obs}} = \SI{1e0}{\rm yr}$.

In our analysis, we account for three different experiments, each representative of current or next-generation detectors in their respective frequency band.
As before, we use ET for the frequency band $\SI{1}{\hertz} -  \SI{10}{\kilo\hertz}$, and LISA for the frequency band $\SI{10}{\micro\hertz} -   \SI{1}{\hertz}$~\cite{ET_2010, LISA_2017, LISA_2019}.
Different from our previous analysis of the PAI model, we find that GW signals can have a sizable amplitude in the PTA frequency band, $\SI{1}{\nano\hertz} - \SI{100} {\nano\hertz}$. 
Correspondingly, we also estimate the potential for a GFIGW observation by the NANOGrav collaboration in their 15-year data set~\cite{agazie_nanograv_2023}, henceforth referred to as NG15.
To compute the SNR for this data set, we work with the total observing time span covered by it, $t_{\rm obs}=\SI{16.03}{\rm yrs}$.
Additionally, we consider any sensitivity by the LIGO--Virgo network as a constraint, given that it did not observe an SGWB in its third observing run~\cite{LIGO_collaboration_2021}.
We shall refer to these constraints as HLVO3 corresponding to Hanford (H), Livingston (L), and Virgo (V). These detectors cover a frequency band $\SI{10}{\hertz} - \SI{50}{\kilo\hertz}$~\cite{LIGO_2010,LIGO_2014,VIRGO_2014}.
For ET and LISA, we use the strain noise spectra $\Omega_{\rm noise}$ collected in Refs.~\cite{schmitz_2021}. For the HLVO3 constraints, we work with the characteristic HLV strain noise spectra available at Ref.~\cite{LIGOScientificCollaboration2021GWTC3}. For NG15, we take the noise spectrum $\Omega_{\rm noise}$ provided by the NANOGrav collaboration~\cite{agazie_nanograv_2023_noise,nanograv_2023_nose_data}. This collection of noise spectra also allows us to compute power-law-integrated sensitivity (PLIS) curves~\cite{Thrane:2013oya} for HLVO3, ET, LISA, and NG15, which we show in Figs.~\ref{fig: spectra, inst. reh}, \ref{fig: spectra, with reh}, and \ref{fig: Schwinger BR - kS}. 

We solve the dynamics of inflation using the gradient expansion formalism (GEF). For more details on the GEF, see Appendix~A of Part I~\cite{vonEckardstein_PAI_2025} and our previous works on FAI~\cite{gorbar_2023, eckardstein_2025}.
We initialize the system on the slow-roll attractor, $\varphi(0) = 15.55 \MP$ and $\dot{\varphi}(0) = - \sqrt{2 /3} \,m \MP$ assuming zero gauge-field and fermion energy densities, $\rho_{\rm EM}(0) = \rho_{\chi}(0)=0$.
The SGWB spectra are computed based on the background evolution determined by the GEF. First, we determine the gauge-field spectra by solving Eq.~\eqref{eq: Mode Eq - FAI}.
Then, we use these mode functions to compute the induced tensor power spectrum, Eq.~\eqref{eq: PT induced}.
This computation is performed using our newly available \emph{Gradient Expansion Formalism Factory (GEFF)}~\cite{vonEckardstein:2025jug}, available at \href{https://github.com/richard-von-eckardstein/GEFF}{https://github.com/richard-von-eckardstein/GEFF}.
More details on this computation are also given in Part I, Appendix B.

To completely determine the modeling of FAI in our effective scheme, we need to define the heuristic scale dependence $\Theta(t,k)$ of the induced current ${\bm J}$ in Eq.~\eqref{eq: ScaleDep Ohmic current}.
In Ref.~\cite{eckardstein_2025}, we presented two particular choices, one which is physically well motivated, and an approximation of the same that allows for easier numerical implementation.
In Sec.~\ref{sec: Results}, we explore the SGWB production in the parameter space spanned by our model parameters $\beta$ and $m$. To make such a scan feasible, we opt for the simpler, approximate implementation of $\Theta(t,k)$.
We discuss both the modeling of $\Theta(t,k)$, as well as the impact of this approximate model in Appendix~\ref{app: scale-dep modeling}. We find that the approximate modeling applied in the main body of the text does not affect our conclusions.
%


\section{Fermion-tempered backreaction}
\label{sec: Schwinger BR}

Before we delve into our main results, i.e., estimating the detectability of GFIGWs from \FAISM,
we want to comment on a novel dynamical regime of FAI, which we dub \textit{fermion-tempered backreaction}.
As the name suggests, the mechanism behind fermion-tempered backreaction is linked to the damping of the gauge field by the conductive medium. 

To understand the new backreaction regime, we first need to consider the usual backreaction during PAI.
It typically proceeds via the gauge-friction term $I_{,\phi}\langle \bm{E} \cdot \bm{B}\rangle$ becoming large compared to Hubble friction, as quantified by the backreaction parameter
\begin{equation}
\label{eq:deltaKG}
    \delta_{\mathrm{KG}} \equiv \frac{|I_{,\phi}(\varphi) \langle \bm{E} \cdot \bm{B} \rangle |}{ |3 H \dot{\varphi}| } \, . 
\end{equation}
As this friction reduces the inflaton velocity, which itself is responsible for gauge-field production,
the entire system enters an oscillatory stage of deceleration and acceleration of the axion field with a retarded response by the gauge field to the change in velocity~\cite{Domcke_2020_Resonant,eckardstein_2023}.
During this process, the kinetic energy in the inflaton is reduced, and the energy density stored in the gauge field ends inflation by becoming dominant enough to stop the accelerated expansion.

Now consider the added impact of fermions. Once the gauge-friction term becomes relevant, the system will attempt to enter strong backreaction.
However, upon completing the first cycle of oscillations in $\dot{\varphi}$, gauge-field production is enhanced, but so are the conductivity terms $\sigma_{E/B}$.
Their effect is to dampen the gauge field, thus limiting the amount of friction generated for the axion.
This stabilizes the oscillations of the system such that the kinetic energy is not completely drained from the axion. Instead, it performs an oscillatory motion around the original slow-roll
trajectory, which increases the duration of inflation due to this stuttering motion, but without an explosive production of gauge bosons.
\begin{figure}
    \centering
    \includegraphics[width=0.98\textwidth]{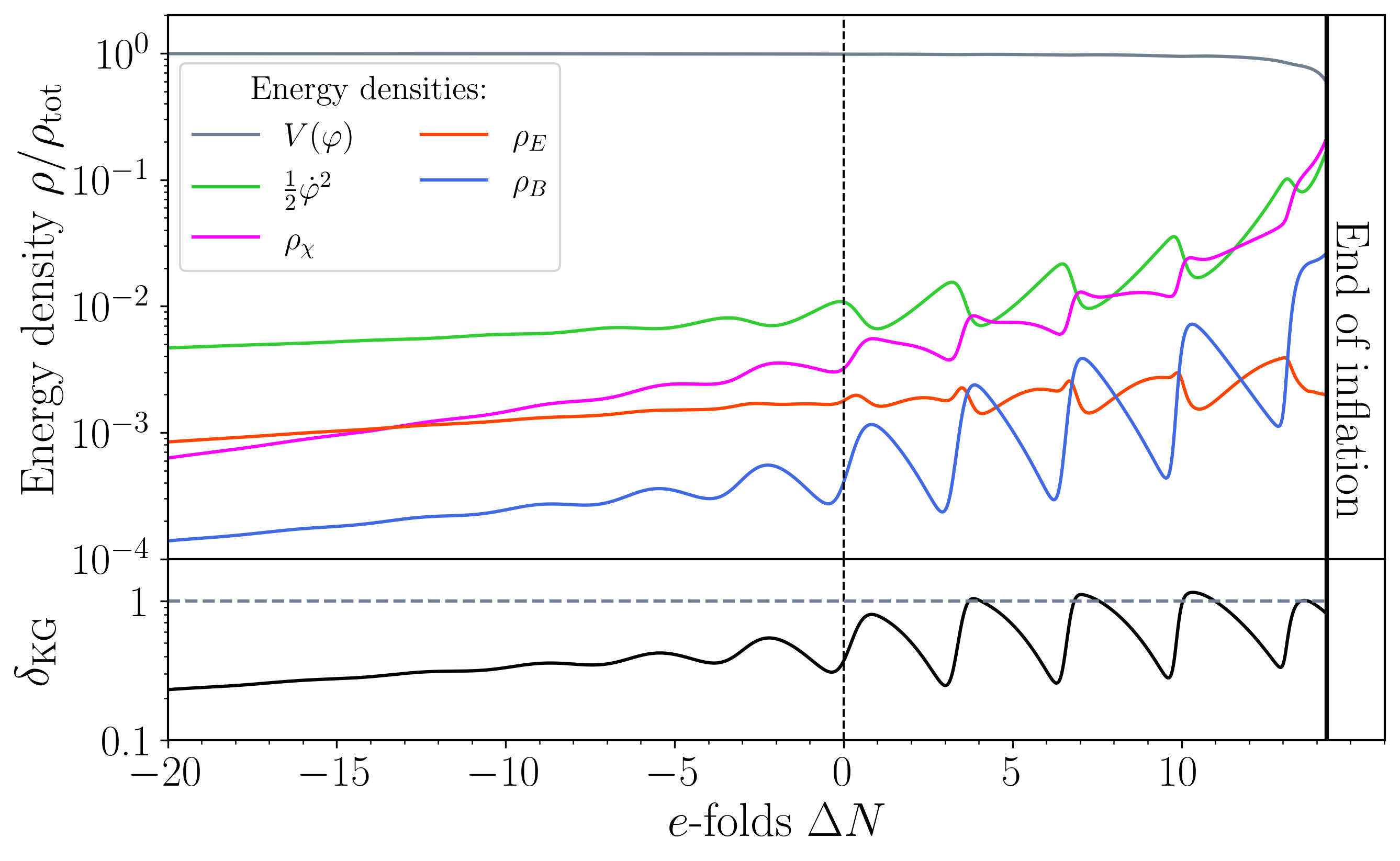}
    \caption
    {
       The evolution of \FAISM \ for $\beta=59$ and $m=\num{2.5e-5} \, \MP$.
       The upper panel shows the evolution of the energy densities $V(\varphi)$ (gray), $\frac{1}{2}\dot{\varphi^2}$ (green), $\rho_{E}$ (red), $\rho_{B}$ (blue), and $\rho_\chi$ (magenta)
       normalized to the total energy density $\rho_{\rm tot} = 3H^2 \MP^2$. The lower panel shows the corresponding evolution of the backreaction parameter $\delta_{\rm KG}$ defined in Eq.~\eqref{eq:deltaKG}.
       The horizontal axis indicates $e$-folds with respect to the expected end of slow-roll inflation at $\Delta N=0$.
       The evolution of this system is representative for fermion-tempered backreaction.
     }
    \label{fig: Schwinger BR}
\end{figure}
An example of these dynamics is given in Fig.~\ref{fig: Schwinger BR}. There, we show the evolution of the energy densities $\rho_i$ for the individual fluids in \FAISM \ for $\beta=59$ and $m=\num{2.5e-5} \, \MP$.
Initially, the inflaton velocity starts feeling $\langle \bm{E} \cdot  \bm{B} \rangle$, inducing small oscillations due to the retarded response between the two fluids.
However, every local maximum in the gauge-field abundance is mimicked by a rise in fermion production, as seen by comparing $\rho_{\chi}$ to $\rho_{E}$ and $\rho_{B}$. 
The damping of $\rho_{E}$ and $\rho_{B}$ due to these fermions adds to the already reduced production of gauge quanta following the oscillating inflaton velocity.
This results in a loss of friction for the inflaton, allowing it to accelerate and repeat this cycle. The lower panel in Fig.~\ref{fig: Schwinger BR} supports this interpretation. There, we show the evolution of the backreaction parameter, $\delta_{\rm KG}$.
Clearly, whenever $\delta_{\rm KG}$ becomes close to unity, the friction term is immediately depleted, leading to temperate oscillations in all fluids.


\section{Gravitational waves from FAI}
\label{sec: Results}

To examine the detection prospects of GFIGWs from \FAISM, we perform a linear parameter scan in $\beta$ and $\log_{10}\frac{m}{\MP}$.
Unlike the analysis of PAI in Part I, we find no need to perform this parameter scan in rotated coordinates.
The reason for rotating the parameter space in the first analysis was that, in the PAI model, the strength of gauge-field backreaction scales as $\sim \frac{m}{\MP} 10^{\kappa\beta}$ with some coefficient $\kappa>1$. 
Thus, an increase in $\beta$ needs to be compensated by a decrease in $\log_{10}\frac{m}{\MP}$ according to the slope $\kappa$, which we determined heuristically in Part I.
However, these considerations no longer hold for FAI since the overall gauge-field amplitude is suppressed as a consequence of fermion production. Hence, also backreaction is inhibited, altering the direct dependence between $\beta$ and $m$.
In fact, the well-known analytical formula for the induced tensor power spectrum of PAI, $\mathcal{P}_{T, \lambda}^{\rm ind} \propto H^4 \exp{(4 \pi \xi)}$, no longer applies to FAI due to this dampened production.
The generic scaling of $\mathcal{P}_{T, \lambda}^{\rm ind}$ with $H^4 \sim m^4$ still holds, however, as it is a consequence of Eq.~\eqref{eq: PT induced}, and is independent of the precise mode functions $A_\lambda(t,k)$.
Conversely, the instability parameter $\xi$ is inhibited by gauge-field production (see Sec.~\ref{sec: FAI}).
Consequently, we expect that increasing the inflaton mass enhances $\mathcal{P}_{T, \lambda}^{\rm ind}$, while an increase of $\beta$ has a reduced effect compared to the PAI case.  

Following these considerations, the results in this section are based on a parameter scan over inflaton masses ranging between $10^{-7.5}-10^{-5.5} \, \MP$  and $\beta$ between $20 - 60$.

\subsection{Instantaneous reheating}
\label{subsec: Results, inst. RH}

\begin{figure}
    \centering
    \includegraphics[width=0.95\textwidth]{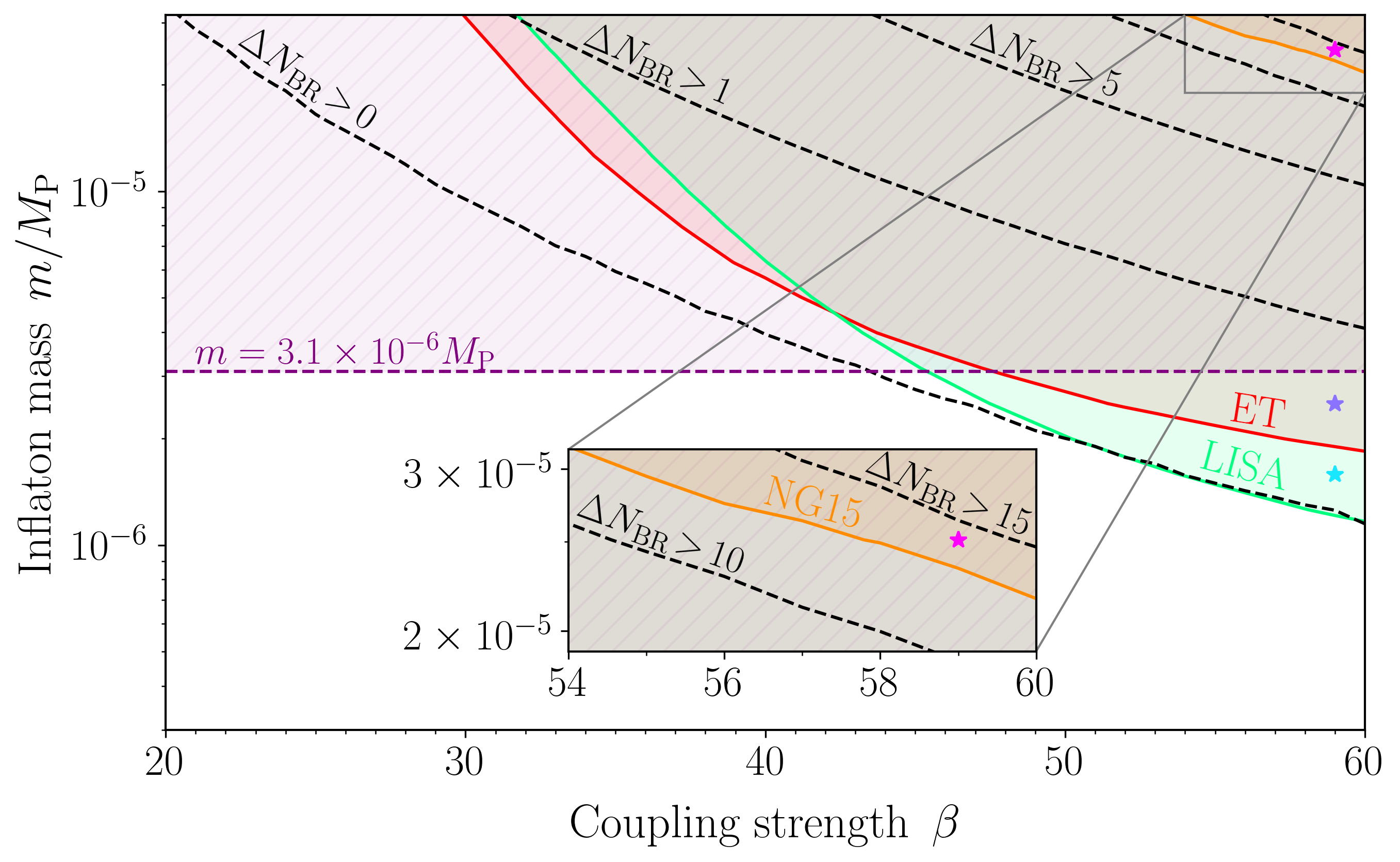}
    \caption
    {
    Regions with $S/N > 1$ for ET (red), LISA (green), and NG15 (orange) and instantaneous reheating.
    PLANCK constraints on the inflaton mass are shown as a purple shaded region.
    In dashed black, we show contour lines for $\Delta N_{\rm BR}$, the prolonged duration of inflation due to gauge-field backreaction.
    The benchmark points marked in cyan, violet, and magenta correspond to the GW spectra shown in Fig~\ref{fig: spectra, inst. reh}.
    For $\beta \gtrsim 45$, we find a viable region for a detection by ET or LISA, while an FAI interpretation of the NG15 signal appears to be excluded by PLANCK.
    }
    \label{fig: SNR, inst. reh}
\end{figure}

Let us first focus on the observational prospects for detecting GFIGWs from ${\rm FAI}_{\rm SM}$ assuming instantaneous reheating. 
For a discussion of a lowered reheating temperature, $T_{\rm reh}$, see Sec.~\ref{subsec: Results, variation of Treh}. Our findings are summarized in Fig.~\ref{fig: SNR, inst. reh}, where we indicate regions with $S/N > 1$ for ET, LISA, and NG15 in red, green, and orange, respectively, together with $m \lesssim \num{3.1e-6}\MP$, the CMB bound, as a purple shaded region.
Evidently, for $\beta \gtrsim 30$, \FAISM \ supports a sufficiently strong gauge field such that the resulting GWs would be detectable by both LISA and ET. In fact, we find a large overlap between regions of $S/N>1$ for ET and LISA,
indicating that the GW amplitude spans a large frequency range. For $\beta \gtrsim 54$ and $m \gtrsim \num{2e-5} \MP$, the signal can even extend down to the nanohertz regime,
such that it is potentially observed by NG15.
However, to support large enough amplitudes in $\Omega_{\rm GW}$, the inflaton mass needs to be sufficiently high.
Hence, a large part of the LISA and ET region of interest is in tension with the CMB bound on $m$, but a small viable region still remains for $\beta \gtrsim 45$. 
The region of interest for NG15 is clearly excluded by this bound.
In contrast to the CMB bounds, we find that neither $\Delta N_{\rm eff}$ nor HLVO3 impose any constraint in the region of parameter space that we analyze.

Alongside the SNR regions in Fig~\ref{fig: SNR, inst. reh}, we also indicate the relevance of gauge-field backreaction by showing contour lines for the extended duration of inflation due to backreaction, $\Delta N_{\rm BR}$.
As expected, larger masses and couplings increase the relevance of backreaction. However, in stark contrast to our findings for PAI, strong backreaction is not required to reach $S/N > 1$.
This is clearly a result of efficient fermion damping prohibiting backreaction even when increasing the inflaton mass.
However, even when backreaction effects do arise, their impact during FAI is manifestly different.
For PAI, we found that strong backreaction inevitably leads to an overproduction of GWs in tension with $\Delta N_{\rm eff} \lesssim 0.5$.
However, these constraints are entirely irrelevant for \FAISM, even for an extremely increased duration of inflation, $\Delta N_{\rm BR} \gtrsim 15$. 
This is a consequence of the fermion-tempered backreaction that we discussed in Sec.~\ref{sec: Schwinger BR}.
It allows for a delayed end of inflation without being accompanied by a drastic production of gauge bosons.

The region of interest for NG15 notably coincides with $\Delta N_{\rm BR} \gtrsim 10$. Given that the CMB bound, Eq.~\eqref{eq: CMB bounds}, depends on $\Delta N_{\rm BR}$,
one may wonder if this backreaction sufficiently weakens the constraints on $m$, such that the NG15 region of interest would become viable. 
However, even for $\Delta N_{\rm BR} = 20$, the bound is still far too constraining; $m \lesssim \num{4e-6} \MP$.
\begin{figure}
    \centering
    \includegraphics[width=0.95\textwidth]{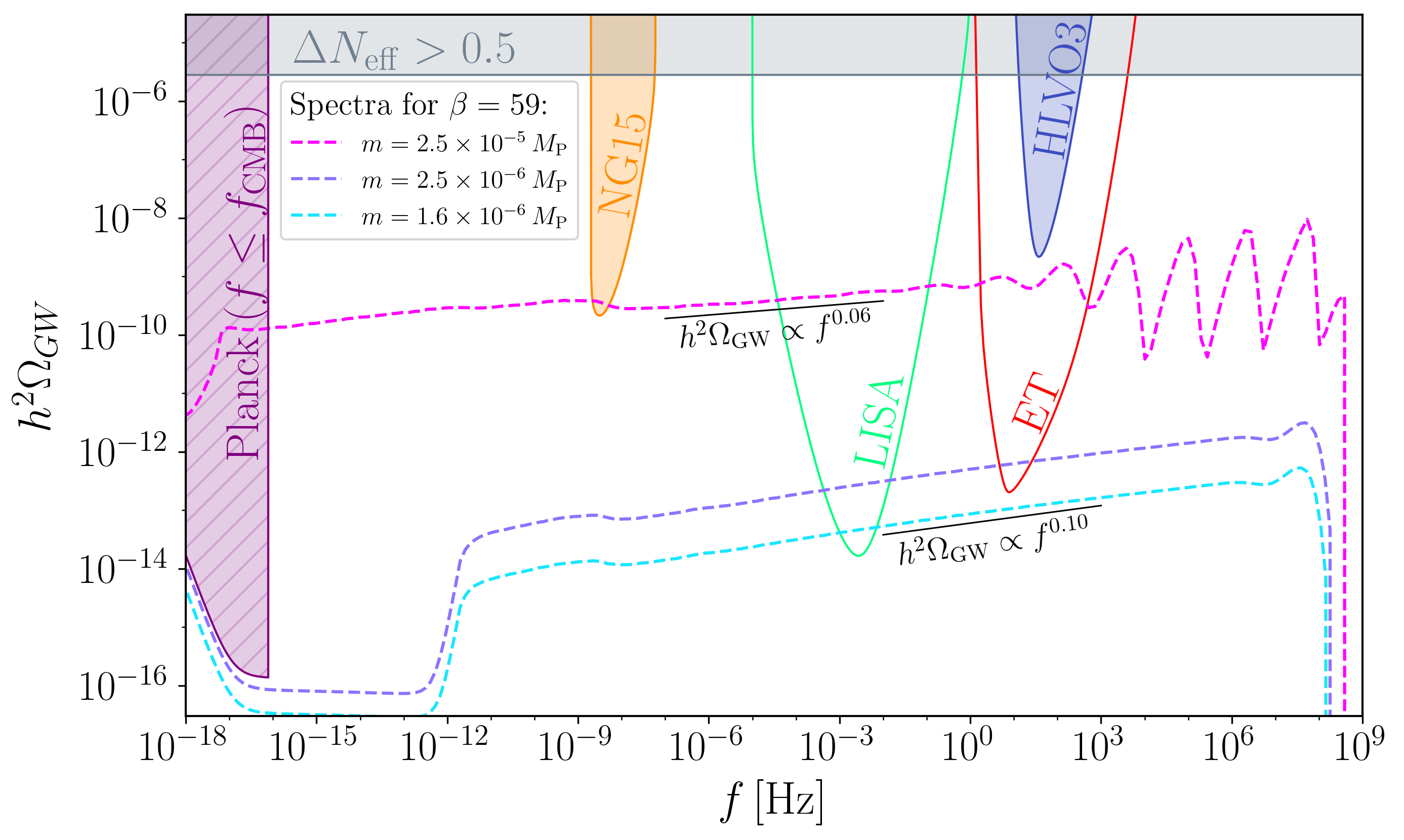}
    \caption
    {
        GW spectra for the benchmark points $\beta = 59$ with $m=\num{1.5e-6} \MP$ (cyan), $m=\num{2.5e-6} \MP$ (violet), and $m=\num{2.5e-5} \MP$ (magenta).
        PLIS curves are shown for ET, LISA, and NG15 with the same colors as in Fig.~\ref{fig: SNR, inst. reh}. In addition,
        we indicate constraints from HLVO3 in blue and from PLANCK (see text for more details) in purple .
        Clearly, all spectra avoid $\Delta N_{\rm eff}$ constraints. However, the magenta spectrum is in clear conflict with the PLANCK observations.
     }
    \label{fig: spectra, inst. reh}
\end{figure}
In Fig~\ref{fig: SNR, inst. reh}, we highlight three benchmark points for $\beta = 59$ with $m=\num{1.5e-6} \MP$, $\num{2.5e-6} \MP$ and $\num{2.5e-5} \MP$ in cyan, violet, and magenta, respectively.
The full GW spectra for these points are presented in Fig.~\ref{fig: spectra, inst. reh} alongside the PLIS curves for ET, LISA, and NG15.
In addition, we show constraints from HLVO3 in terms of its PLIS curve. Constraints from PLANCK on the tensor-to-scalar ratio are also indicated. We do so in a particular way that requires further explanation: As discussed in Sec.~\ref{sec: Parameters and Constraints}, we only demand that the inflaton potential be of the form $V(\phi) = m^2 \phi^2 /2$ for field values below some threshold value, $\varphi \leq \varphi_{\rm thr}$, while at field values above this threshold, the inflaton potential is free to receive model-dependent corrections. The upper limits on $V(\varphi_{\rm thr})$ in Eq.~\eqref{eq: CMB bounds} notably apply to this more flexible model, in which $V(\phi) = m^2 \phi^2 /2$ for $\varphi \leq \varphi_{\rm thr}$ and $V(\phi) = V_{\rm CMB}(\phi)$ for $\varphi > \varphi_{\rm thr}$, where $V_{\rm CMB}(\phi)$ represents a monotonically increasing continuation of the quadratic potential at small values, but which is otherwise arbitrary. In Fig.~\ref{fig: spectra, inst. reh}, on the other hand, we need to make a choice for $V_{\rm CMB}(\phi)$. For definiteness, we simply assume that, for the purposes of drawing GW spectra at all frequencies in Fig.~\ref{fig: spectra, inst. reh}, the potential is of the form $V(\phi) = m^2 \phi^2 /2$ at all field values, including field values above $\varphi_{\rm thr}$ (frequencies below $f_{\rm thr}$). This particular model may be referred to as the $m^2 \phi^2$-equivalent of the actual, more flexible model that we assume in our discussion of parameter space in Fig~\ref{fig: SNR, inst. reh}. The CMB constraints in Fig.~\ref{fig: spectra, inst. reh} therefore should be interpreted as follows:
if an $m^2 \phi^2$-equivalent GW spectrum does or does not violate the CMB bound shown in Fig.~\ref{fig: spectra, inst. reh}, the corresponding more flexible model does or does not violate the observational constraints on $A_s$, $r$, and $\left|\xi_{\rm CMB}\right|$, respectively. This construction has several consequences. First of all, the CMB limit in Fig.~\ref{fig: spectra, inst. reh} is not a model-independent sensitivity curve on the same footing as the PLIS curves in the plot. By construction, it represents a specific constraint that only applies to the GW spectra in the $m^2 \phi^2$-equivalent model. Second, these $m^2 \phi^2$-equivalent GW spectra do not need to satisfy the observational limit on the amplitude of the tensor power spectrum, $rA_S \lesssim 0.03 \times 2.1 \times 10^{-9}$; this bound applies to the GW spectra in our actual, more flexible model, in which $V_{\rm CMB}(\phi)$ is left unspecified. To determine the upper (``would-be'') limit on $rA_S$ in the $m^2 \phi^2$-equivalent model, we work with Eq.~\eqref{eq: CMB bounds} together with Eq.~\eqref{eq: thr potential} evaluated for instantaneous reheating and $f_{\rm thr}=\SI{e-12}{\hertz}$.
The resulting excluded region is then constructed by assuming a flat tensor power spectrum up to $f_{\rm CMB} = \SI{7.7e-17}{\hertz}$, which we plug into Eq.~\eqref{eq: omega-GW from PT}.

Comparing Fig.~\ref{fig: spectra, inst. reh} to its counterpart In Part I~\cite{vonEckardstein_PAI_2025}, Fig.~3, confirms our previous description of the differences between PAI and FAI.
For PAI, we found peaked spectra with a strongly blue tilted GW spectrum, surpassing the vacuum contribution by many orders of magnitude. 
For \FAISM \ on the other hand, GFIGWs are abundant enough to be observable by ET, LISA, or even NG15, but never in threat of violating bounds on $\Delta N_{\rm eff}$. 
Instead of the drastic amplification of GWs due to strong backreaction, a temperate amplitude in GFIGWs is sustained over many orders of magnitude in frequency.
A first initial burst in gauge-field production lifts the spectrum above the vacuum contribution, where it then remains with only a mild blue tilt.
This is clearly a consequence of the dampened efficiency of gauge-field production due to Schwinger pair creation.

The two lower-mass spectra, $m=\num{1.5e-6} \MP$ and $\num{2.5e-6} \MP$, are nearly identical, except for the increase in amplitude for the higher-mass spectrum.
For $f \gtrsim \SI{e-11}{\hertz}$, the spectrum rises with a mild blue tilt, $h^2\Omega_{\rm GW} \propto f^{0.1}$ (except for a small decrease due to the change in $g_{*,S}$ around the quantum chromodynamics (QCD) crossover at  $f\sim \SI{e-9}{\hertz}$).
The spectra not only reach the sensitivity curves for ET and LISA, but are also safe regarding CMB constraints at $f\sim f_{\rm CMB}$. The GFIGW contribution only becomes important at $f \sim \SI{e-12}{\hertz}$
and the vacuum contribution is not in tension with the upper limit on the tensor-to-scalar ratio, $r \lesssim 0.03$, as indicated by the purple shaded region in Fig.~\ref{fig: SNR, inst. reh}.

The spectrum at high mass, $m=\num{2.5e-5} \MP$, corresponds to the evolution of the energy density shown in Fig.~\ref{fig: Schwinger BR}.
It evidently is the product of a background evolution with significant gauge-field backreaction.
This observation allows us to explain all the apparent differences between this spectrum and the ones for $m=\num{1.5e-6} \MP$ and $\num{2.5e-6} \MP$.
Firstly, note the drastic increase to the duration of inflation, $\Delta N_{\rm BR} \simeq 14$, as inferred from Figs.~\ref{fig: Schwinger BR} and~\ref{fig: SNR, inst. reh}.
It implies that the GW frequency is redshifted by a factor of $10^5$, explaining why the rise in this GW spectrum happens at much lower frequencies than for the other two spectra.
Secondly, we can see novel oscillations in $h^2\Omega_{\rm GW}$ at high frequencies mimicking those of the gauge-field energy densities in Fig.~\ref{fig: Schwinger BR}.

The rise in the GFIGW amplitude for $m = \num{2.5e-5} \MP$ around $f_{\rm CMB}$ clearly is in conflict with PLANCK constraints. 
However, we want to stress that this parameter point was simulated assuming a perfectly quadratic inflaton potential at all field values.
If we were to flatten the inflaton potential at early times, one would expect gauge-field production to set in later due to the lower inflaton velocity on the flatter sections. 
Consequently, the production of GFIGWs would be observed at higher frequencies. 
Of course, this would not save the GW spectrum presented here from being ruled out, as even the vacuum contribution for this spectrum is too high, which can be both read off from Fig.~\ref{fig: SNR, inst. reh} and Fig.~\ref{fig: spectra, inst. reh}, remembering our special construction of the CMB exclusion contour in Fig.~\ref{fig: spectra, inst. reh}. 
Still, this model dependence is worth remembering when viewing these GW spectra.

However, even when disregarding PLANCK constraints, the spectrum for $m=\num{2.5e-5} \MP$ would still not be suitable for explaining the NG15 data, although it reaches a sufficient amplitude. The NG15 data favors a steeper spectral index, $h^2 \Omega_{\rm GW} \propto f^{5 - \gamma}$, with $\gamma = 3.2 \pm 0.6$ (median and $90\%$ credible interval)~\cite{agazie_nanograv_2023}. Clearly, the spectrum we show falls short of this spectral slope with  $h^2 \Omega_{\rm GW} \propto f^{0.06}$ around the relevant frequencies. In fact, the background slope of our spectrum is so small that the changes in the effective number of relativistic degrees of freedom around the QCD phase transition even lead to momentary drops of the spectral index below zero (i.e., $\gamma > 0$), which is disfavored by the NG15 data.    

In summary, we find that, while PAI is primarily constrained by $\Delta N_{\rm eff}$ due to the sudden and drastic transition into the strong-backreaction regime,
this is no longer true for FAI. The additional friction from Schwinger pair production tempers these violent processes, allowing for observable signals by both ET and LISA.
However, these same damping effects also imply that a larger Hubble rate at production is required in order to supply the necessary amplitude in GWs.
Therefore, we find that the most relevant constraints on the SGWB from FAI come from the PLANCK measurements of $A_S$ and the limit on $r$, in contrast to PAI, where the most stringent constraints came from $\Delta N_{\rm eff}$.

\subsection{Varying the reheating temperature}
\label{subsec: Results, variation of Treh}

To elaborate on our findings in the previous section, we include a simple modeling of reheating effects, assuming a reheating phase of early matter domination, i.e., $w_{\rm reh}=0$.
We treat the reheating temperature, $T_{\rm reh}$, as a free parameter, which we can lower down to $T_{\rm reh} \sim T_{_{\rm BBN}} \sim \SI{1}{\mega\electronvolt}$. Of course, this is a generous assumption
and the precise initial conditions present at the end of inflation are relevant here. For certain parameter regions of \FAISM, especially for high couplings $\beta$ and masses $m$,
we find that a large fraction of energy is already stored in SM fermions and $U(1)_{\rm Y}$ gauge bosons. Therefore, one would assume that the duration of reheating may not be an arbitrary parameter.
However, for illustrating the effect of reheating, it is easiest to treat all parameter points on equal footing, instead of setting a lower threshold on $T_{\rm reh}$ on a point-by-point basis.
\begin{figure}
    \centering
    \includegraphics[width=0.98\textwidth]{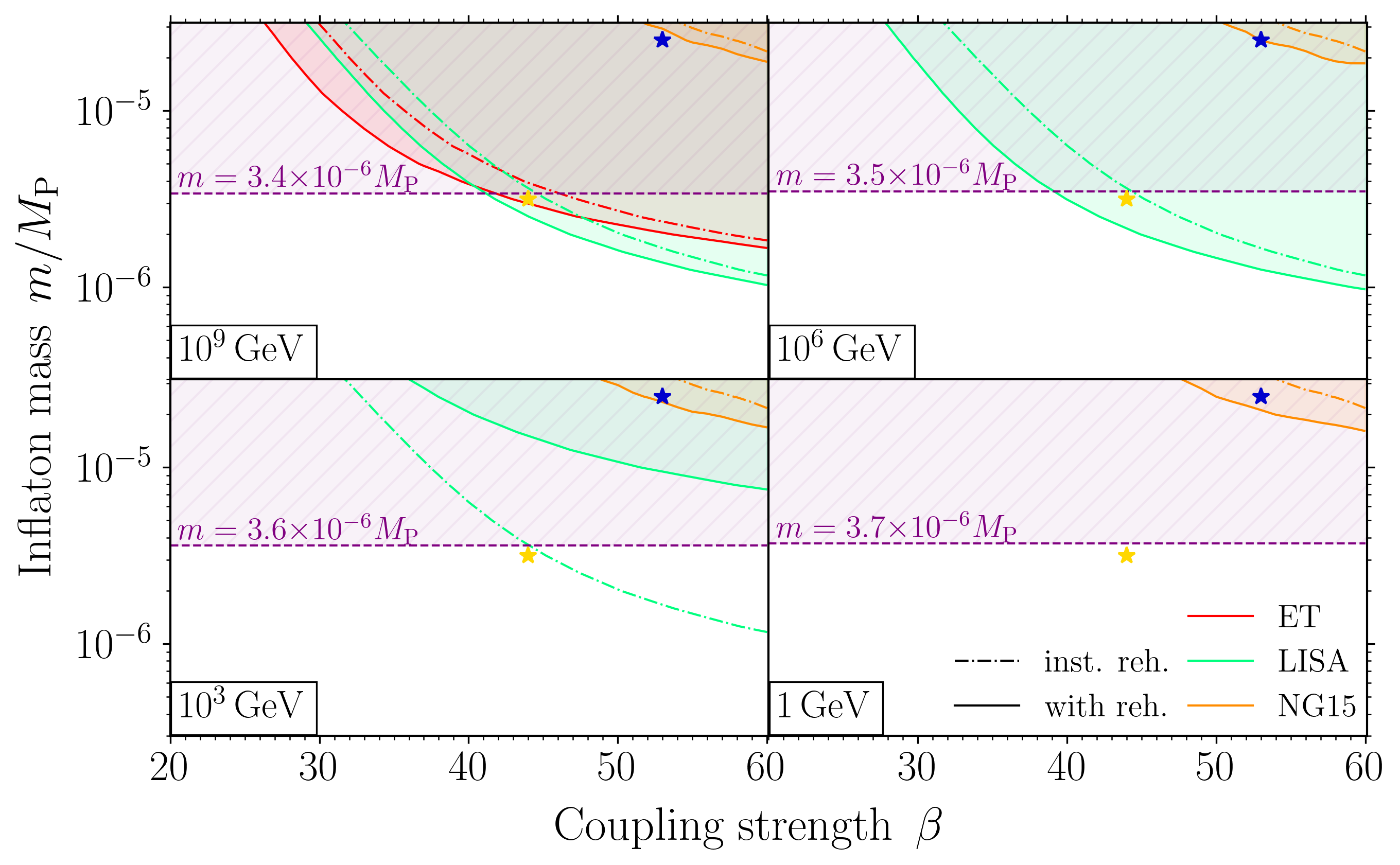}
    \caption
    {
    Regions with $S/N>1$ for ET (red), LISA (green), and NG15 (orange) including variations in the reheating temperature, $T_{\rm reh}$.
    From left to right and top to bottom, $T_{\rm reh}$ is $\SI{e9}{\giga \electronvolt}$, $\SI{e6}{\giga \electronvolt}$, $\SI{e3}{\giga \electronvolt}$, and $\SI{1}{\giga \electronvolt}$ as indicated in the lower left corner of each panel.
    For reference, the contours of $S/N>1$ assuming instantaneous reheating (i.e., the ones in Fig.~\ref{fig: SNR, inst. reh}) are shown as dashed-dotted lines.
    A lower $T_{\rm reh}$ first increases the sensitivity of a detector, until it falls below the temperature corresponding to the frequency band of the detector.
    For points marked in yellow, $\beta = 44$, $m=\num{3.2e-6}\MP$, and blue, $\beta = 53$, $m=\num{2.5e-5}\MP$, we show $h^2 \Omega_{\rm GW}(f)$ in Fig.~\ref{fig: spectra, with reh}.
     }
    \label{fig: SNR, varying Treh}
\end{figure}
In Fig.~\ref{fig: SNR, varying Treh}, we show the change in the $S/N > 1$ regions for LISA, ET, and NG15 when lowering the reheating temperature to $T_{\rm reh} = \SI{e9}{\giga \electronvolt}$, $\SI{e6}{\giga \electronvolt}$, $\SI{e3}{\giga \electronvolt}$, and $\SI{1}{\giga \electronvolt}$.
For comparison, we also show the results of Fig.~\ref{fig: SNR, inst. reh} in dashed-dotted lines. 
Evidently, lowering the reheating temperature enlarges the SNR regions before they vanish entirely.
This is easily understood: lowering $T_{\rm reh}$ corresponds to redshifting the blue-tilted spectrum. Correspondingly, a higher amplitude in $h^2\Omega_{\rm GW}$ can be achieved for lower frequencies.
However, a lowered reheating temperature also implies a drop-off at the frequency corresponding to $T_{\rm reh}$. Therefore, once the reheating temperature drops below temperatures corresponding to the frequency band of a detector,
the detector loses all sensitivity. This happens first for ET, with the lower frequency roughly corresponding to $T \sim \SI{4e7}{\giga \electronvolt}$. Then, LISA loses sensitivity around $T \sim \SI{400}{\giga \electronvolt}$.
For NG15, we do not observe this in Fig.~\ref{fig: SNR, varying Treh} as its critical temperature is $T \sim \SI{100}{\mega \electronvolt}$, i.e., below $T_{\rm reh} = \SI{1}{\giga\electronvolt}$, the lowest temperature we consider in Fig.~\ref{fig: SNR, varying Treh}.
We also indicate the mild drift in the PLANCK limits on the inflaton mass, Eq.~\eqref{eq: CMB bounds}, for $\Delta N_{\rm BR}=0$ as indicated by the respective label.
\begin{figure}
    \centering
    \includegraphics[width=0.95\textwidth]{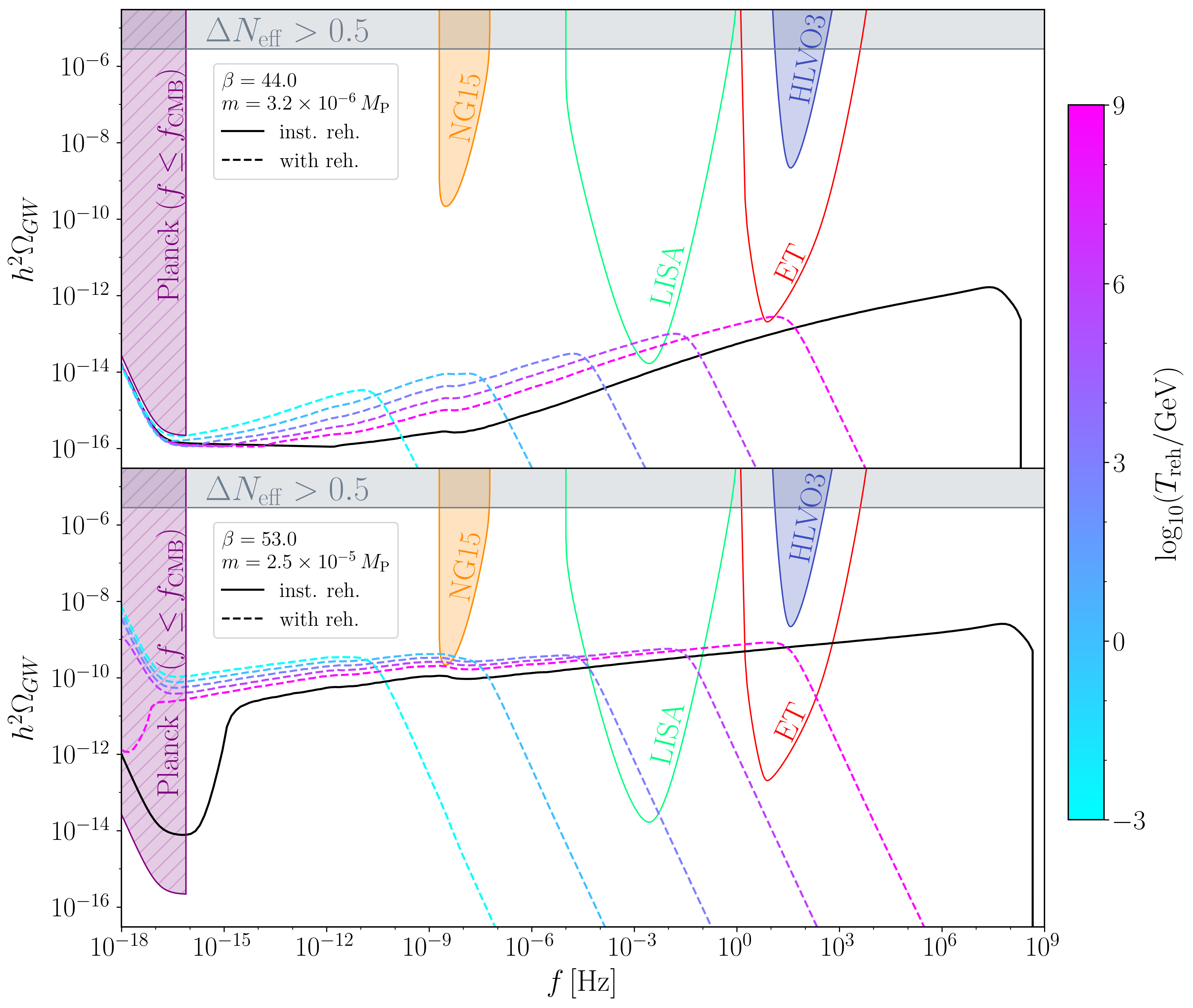}
    \caption
    { 
    GW spectra for $\beta = 44$, $m=\num{3.2e-6}\MP$ (upper panel) and $\beta = 53$, $m=\num{2.5e-5}\MP$ (lower panel) including their variations when lowering the reheating temperature.
    The black solid spectra correspond to instantaneous reheating, while the color of the dashed spectra correspond to $T_{\rm reh} = \SI{e9}{\giga \electronvolt}$, $\SI{e6}{\giga \electronvolt}$, $\SI{e3}{\giga \electronvolt}$, $\SI{1}{\giga \electronvolt}$, and $\SI{1}{\mega \electronvolt}$, respectively; see the color bar on the right.
    PLIS curves and PLANCK constraints are shown with the same color code as in Fig.~\ref{fig: spectra, inst. reh}.
     }
    \label{fig: spectra, with reh}
\end{figure}
To elaborate on these results, we again indicate two qualitatively different benchmark points in Fig.~\ref{fig: SNR, varying Treh}, $\beta = 44$, $m=\num{3.2e-6}\MP$, and $\beta = 53$, $m=\num{2.5e-5}\MP$,
for which we show the effect of reheating on $h^2\Omega_{\rm GW}(f)$ in Fig.~\ref{fig: spectra, with reh}.
For each parameter point, the spectrum for instantaneous reheating is depicted in solid black, while the colored dashed curves show the effect of lowering $T_{\rm reh}$. 
PLIS curves are given exactly as in Fig.~\ref{fig: spectra, inst. reh}. We again compute the PLANCK bound as in Fig.~\ref{fig: spectra, inst. reh}, but assuming the lowest reheating temperature for which we show spectra, $T_{\rm reh} = \SI{1}{\mega \electronvolt}$. The spectra match our expectations: a lower reheating temperature redshifts the spectrum, first increasing the sensitivity of a detector, 
before the spectral peak is redshifted out of its frequency band for a low enough $T_{\rm reh}$.

\subsection{Estimating the fermionic contribution to the SGWB }
\label{subsec: FIGWs}
Up until now we have exclusively considered GWs sourced by gauge bosons. However, Fig.~\ref{fig: Schwinger BR} indicates that more energy density may be stored in fermions than the gauge field at the end of inflation.
If this fermion gas is sufficiently anisotropic, one would expect that it contributes correspondingly to the total amplitude of the GW spectrum.

A computation of this additional contribution, on equal footing to the one for the gauge field, is beyond the scope of this article. 
However, we can make an educated estimate of its relevance,
\begin{equation}
    \Omega_{\rm GW}^{\chi} \sim  \beta_* \left(\frac{\rho_{\chi}}{\rho_{\rm EM}}\right)^2 \Omega_{\rm GW}^{\rm GF} \, .
    \label{eq: GW fermion estimate}
\end{equation}
Here, $\Omega_{\rm GW}^{\chi}$ is the contribution from fermions to the GW spectrum, while $\Omega_{\rm GW}^{\rm GF}$ is that of the gauge field, which we compute from Eq.~\eqref{eq: PT induced}.
We assume that the overall spectral dependence for the fermion contribution matches that of the gauge field, simply relating the two contributions via their respective energy densities.
This reflects the fact that we model the conducting fermionic medium as direct functions of the gauge-field expectatin values, $\langle {\bm E}^2 \rangle$, $\langle {\bm B}^2 \rangle$, and $\langle {\bm E} \cdot {\bm B} \rangle$.
The coefficient  $\beta_*$ describes how inherently anisotropic (quadrupolar and  beyond) the fermion energy density is, i.e., its ability to source GWs.

Next, we try to estimate $\beta_*$ from the microphysics hidden behind $\rho_\chi$.
In Ref.~\cite{domcke_2018}, it was computed that the electric and magnetic field source two distinct fermion populations. 
Firstly, chiral fermions are directly produced from the electric field via the chiral anomaly.
These fermions quickly thermalize due to internal scattering, which we take to imply that they quickly become isotropic and are thus irrelevant for GWs.
However, a second population of left-right symmetric fermions occupies higher Landau levels due to the presence of the magnetic field. For these fermions, scattering is negligible, and they do not thermalize.
Reference~\cite{domcke_2018} estimated that this second population is dominant in $\rho_\chi$, finding that the chiral fermions only make up a small fraction of around $\num{e-2} - \num{e-3}$.
We take this to imply that, in principle, a large fraction of $\rho_\chi$ may be anisotropic. We will thus assume an optimistic value, $\beta_*=1$ for our estimates.
\begin{figure}
    \centering
    \includegraphics[width=0.98\textwidth]{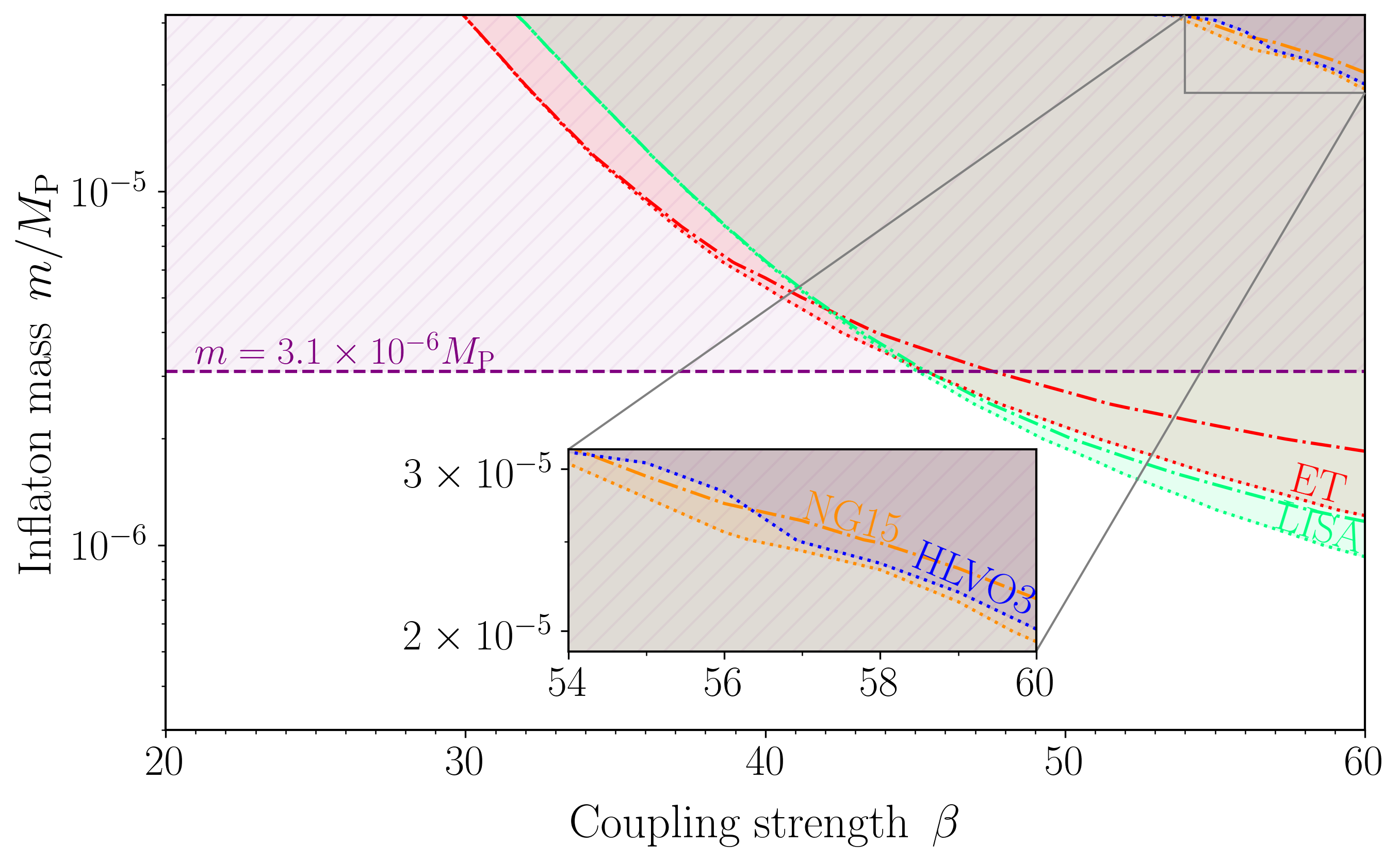}
    \caption
    {
    Sensitivity regions when estimating the additional contribution to $h^2\Omega_{\rm GW}$ due to the fermion gas.
    The regions with $S/N>1$ are shown for ET (red), LISA (green), NG15 (orange), and HLVO3 (blue). These results are computed assuming instantaneous reheating.
    The fermion contribution may boost the detectability region for ET and LISA, thus dominating over the GFIGW contribution at high frequencies.
    At low frequencies, the exra contribution matters little, as evident from the NG15 region being nearly insensitive to the additional source.
    More so, the enhanced signal at high frequencies implies that HLVO3 is in conflict with a FAI interpretation of the NG15 signal.
    These results are a simple estimate assuming an optimistically large anisotropic fermion fraction.
    }
    \label{fig: SNR, fermion estimate}
\end{figure}
In Fig.~\ref{fig: SNR, fermion estimate}, we show the relevance of the fermionic contribution by showing how the SNR-regions for LISA, ET, and NG15 are affected when adding the estimate for $h^2\Omega_{\rm GW}^{\chi}$ from Eq.~\eqref{eq: GW fermion estimate} to $h^2\Omega_{\rm GW}^{\rm GF}$ from Eq.~\eqref{eq: PT induced}. All results are computed assuming instantaneous reheating. 
The $S/N > 1$ regions enveloped by dotted lines indicate the result with $h^2\Omega_{\rm GW}^{\chi}$ included, while the dashed-dotted lines are computed from $h^2\Omega_{\rm GW}^{\rm GF}$ alone.
Besides LISA, ET, and NG15, also HLVO3 now shows sensitivity to $h^2\Omega_{\rm GW}$. 

LISA and ET become considerably more sensitive to the GW signal from FAI, especially for large coupling, $\beta$. The explanation is obvious: stronger gauge-field production yields stronger fermion production.
Interestingly, we find that the sensitivity region for NG15 has not been significantly increased. We attribute this to the observation that fermion production occurs towards the end of inflation, as suggested by Fig.~\ref{fig: Schwinger BR}.
Therefore, at low frequencies, $h^2\Omega_{\rm GW}^{\chi}$ does not dominate over $h^2\Omega_{\rm GW}^{\rm GF}$, and the fermions do not add to the signal in the nanohertz band.
The fact that the fermions contribute primarily at high frequencies also explains the appearance of an $S/N >1$ region for HLVO3: because $h^2\Omega_{\rm GW}^{\chi}$ is more blue-tilted than $h^2\Omega_{\rm GW}^{\rm GF}$,
$h^2\Omega_{\rm GW}$ gains in amplitude at higher frequencies. Based on our simple estimates alone, including $h^2\Omega_{\rm GW}^\chi$ thus further weakens the FAI interpretation of the NG15 signal,
as the HLVO3 region largely overlaps with the one for NG15.

These simple estimates clearly highlight the importance of a more refined computation of the fermionic GW contribution, $h^2\Omega_{\rm GW}^{\chi}$,
seeing as it may be dominant over $\Omega_{\rm GW}^{\rm GF}$ at high frequencies. A precise computation including the spectral information on the fermions is, however, beyond the scope of this analysis.


\section{Conclusion and outlook}
\label{sec: Conclusion}

Axion inflation is a promising model of inflation that attempts to reconcile the flatness of the inflaton potential with non-trivial couplings of the inflaton to gauge fields.
In the simplest models, an Abelian gauge field couples to the inflaton via a shift-symmetric interaction, $\phi F \tilde{F}$.
This results in abundant gauge-field production that may in turn source a stochastic gravitational-wave background (SGWB).
Models of fermionic axion inflation (FAI) extend the field content of axion inflation by charged matter fields.
These charged particles are generated  via the Schwinger effect as a consequence of the strong electric and magnetic fields sourced by the axion.
The resulting conductive medium of charge carriers then dampens the production of the gauge field.

In this article, we investigated the imprint of this non-trivial interplay between the inflaton, the gauge field, and fermions onto the SGWB sourced by the gauge field during axion inflation.
To this end, we considered a particularly well-motivated realization of the FAI model in which the axion couples to the Abelian $U(1)_{\rm Y}$ hypercharge field.
Therefore, the fermions produced by the Schwinger effect are those of the Standard Model (SM).
This article is the second part following a preceding analysis of gravitational-wave (GW) production during pure Abelian axion inflation (PAI)~\cite{vonEckardstein_PAI_2025}.

To estimate the phenomenological viability of the resulting SGWB signal, we  considered its detectability by current and next-generation GW observatories.
This includes experiments spanning a large frequency range, the Laser Interferometer Space Antenna (LISA), the Einstein Telescope (ET) as well as the latest data set of the NANOGrav collaboration (NG15), representative of the sensitivity of pulsar timing array observations.
The constraints on the signal come from the non-observation of an SGWB in the third observing run of the LIGO-Virgo network (HLVO3), PLANCK measurements, and bounds on the effective number of dark relativistic degrees of freedom, $\Delta N_{\rm eff}$.

We studied the production of GWs from FAI assuming a linear coupling between the inflaton and the gauge field, $\beta/\MP \phi F \tilde{F}$, and assuming a simple quadratic potential for the inflaton.
In the two-dimensional parameter space spanned by the inflaton--gauge-field coupling strength, $\beta$, and the inflaton mass, $m$,
we find large regions where an appreciable amount of gauge-field-induced gravitational waves (GFIGWs) may be produced during FAI coupled to the SM.
Especially for couplings $\beta \gtrsim 45$ and inflaton masses below the PLANCK bound, $m \lesssim \num{3.1e-6} \MP$, an SGWB with sufficient amplitude can be generated such that it would be observable by both LISA and ET, thus marking a particular region of interest. To reach the sensitivity range of NG15, larger couplings, $\beta \gtrsim 54$, and larger masses, $m \gtrsim \num{2e-5}\MP$, are required, in tension with existing PLANCK data.
However, even when disregarding the bound on the inflaton mass, the spectral slope of spectra reaching NG15 sensitivity are too shallow to be consistent with NG15; $h^2 \Omega_{\rm GW} \propto f^{0.06}$.

We also analyzed the effect of a lowered reheating temperature, $T_{\rm reh}$, finding qualitative agreement with the results for instantaneous reheating.
Unsurprisingly, when the reheating temperature is lowered below the corresponding frequency range of a detector, the detector loses sensitivity. This occurs for ET at $T_{\rm reh}\sim \SI{4e7}{\giga \electronvolt}$, then for LISA at $T_{\rm reh}\sim\SI{400}{\giga \electronvolt}$ and again for NG15 at $T_{\rm reh} \sim \SI{100}{\mega \electronvolt}$.

Finally, we estimated the additional contribution to the SGWB that is sourced by the SM fermions produced during FAI.
We find that this novel component to the GW spectrum could enhance the signal at high frequencies, as the fermion energy density surpasses that of the gauge field towards the end of inflation.
This implies an increased sensitivity of LISA and ET to GWs from FAI.
At the same time, this additional component of the SGWB puts further stress on the interpretation of the NG15 signal as coming from FAI.
We find that, as the SGWB signal rises more steeply at high frequencies due to the fermionic contribution, most signals that would explain the NG15 data should also have been detected in HLVO3.
Therefore, the region of parameter space with an appreciable signal-to-noise ratio for NG15 is in tension with the non-detection of an SGWB by HLVO3.
These results rely on a simple estimate (see Eq.~\eqref{eq: anisotropic stress - FAI} of the SGWB signal with an optimistic assumption on the anisotropy of the fermionic gas. 
The relevance of the fermionic contribution based on these simple estimates clearly highlights the importance of a proper computation of this novel SGWB component. We leave this for future work.

The results for FAI are in stark contrast to the results for PAI in our companion paper~\cite{vonEckardstein_PAI_2025}.
In our GEF benchmark study, we found that an observable signal for PAI was in tension with bounds on $\Delta N_{\rm eff}$, while a signal from FAI is primarily constrained 
by the maximal inflation scale inferred from the PLANCK data.
The reason for this difference is in the uninhibited production of gauge bosons during PAI, 
which implies that strong backreaction effects can trigger a dynamical instability in the dynamics of the inflaton, 
leading to an explosive production of gauge bosons and GWs.
The damping effect of the fermionic medium in FAI tempers these effects, allowing for the generation of a moderately blue-tilted GW spectrum.
We find that, in the most interesting region of parameter space, the SGWB amplitude grows as $~\propto f^{0.1}$ around the LISA and ET frequency bands.
While this damping does not imply the absence of backreaction, it does moderate it, resulting in a qualitatively new backreaction regime that we call fermion-tempered backreaction.

Both of our analyses were performed using the gradient expansion formalism (GEF). In our previous paper, we discussed at length the limitations of this method when 
studying the dynamical evolution of PAI. These limitations concern the generation of axion gradients sourced by the gauge field. The effect of this additional dynamical degree of freedom has so far only been successfully captured on the lattice.
However, the dampened production of gauge bosons during FAI implies that this effect could be less significant, as recently indicated by lattice simulations~\cite{iarygina_2025}.
Therefore, the GEF appears to be ideally suited for studying the dynamics of FAI.

In summary, in this article, we studied the prospects of detecting GWs originating from the production of SM particles during axion inflation.
We find that a GW signal sourced by $U(1)_{\rm Y}$ hypercharge bosons generated via their coupling to the axion--inflaton field would be detectable by both LISA and ET. We also estimated, for the first time, the additional 
contribution to the GW spectrum sourced by the fermions produced during FAI, finding it to make a relevant contribution at high frequencies.

We hope that our results are going to encourage other researchers to further pursue the study of fermion generation in the early Universe, in particular during axion inflation.
Our results highlight the importance of accounting for the non-trivial interplay between the fermions and the gauge field in this model to make accurate phenomenological predictions.
The fact that an SGWB sourced by SM fields during inflation may be detected by next-generation GW observatories underpins the relevance of improving various aspects of the computation that we have performed here.
First, one should strive towards computing fermion backreaction onto the gauge field from first principles in place of the effective modeling that we applied in this article.
Second, the importance of axion inhomogeneities should be reassessed for the model of FAI to ensure that their impact can indeed be neglected. 
In particular, we hope our results will spark the interest of researchers to validate this assumption using lattice techniques.
Third, in this article, our focus has only been on the gauge field as a source for an SGWB. We already pointed out that the fermions themselves could give a relevant contribution to the SGWB.
However, these same fields can also source scalar metric perturbations, which in turn can induce an additional SGWB component at second order in perturbation theory.
This third major source could then further affect the spectra that we have shown here. 
At the present stage, we are not aware of any computations of the scalar power spectrum for FAI, in particular not one that also accounts for fermions as a source of density fluctuations. 
Overall, FAI is a rich research field with the potential of many important improvements to be made in future work.


\vskip.25cm
\section*{Acknowledgements}
K.\,S.\ is an affiliate member of the Kavli Institute for the Physics and Mathematics of the Universe (Kavli IPMU) at the University of Tokyo and as such supported by the World Premier International Research Center Initiative (WPI), MEXT, Japan (Kavli IPMU). The work of O.\,S.\ has received funding through the SAFE\,---\,Supporting At-Risk Researchers with Fellowships in Europe project, which is funded by the European Union. Views and opinions expressed are, however, those of the authors only and do not necessarily reflect those of the European Union, the European Research Executive Agency (REA). Neither the European Union nor the granting authority can be held responsible for them. Part of this work was conducted using the High Performance Computing Cluster PALMA II at the University of M\"unster~(\url{https://www.uni-muenster.de/IT/HPC}).

\appendix
\section{Alternative modeling of fermionic scale dependence}
\label{app: scale-dep modeling}

In this appendix, we discuss the modeling of the function $\Theta(t,k)$ in Eq.~\eqref{eq: ScaleDep Ohmic current} by means of which we account for the scale dependence of Schwinger damping in the mode equations for the gauge field. We compare two models suggested in Ref.~\cite{eckardstein_2025}, 
which we call \textit{$k_{\rm h}$-damping} and \textit{$k_{\rm S}$-damping}, respectively. In the main body of the text, we relied on $k_{\rm h}$-damping, even though this model is an approximation of the more well-motivated $k_{\rm S}$-damping model. In the following, we will first introduce both models and then discuss how our results in the main text change when replacing $k_{\rm h}$-damping by $k_{\rm S}$-damping on the basis of the benchmark points shown in Fig.~\ref{fig: spectra, inst. reh}, i.e., $\beta = 59$ and $m=\num{1.5e-6} \MP$, $\num{2.5e-6} \MP$, and $\num{2.5e-5} \MP$.

\subsection[Scale-dependent damping with \texorpdfstring{$k_{\rm h}$}{kh} or \texorpdfstring{$k_{\rm S}$}{kS}]{Scale-dependent damping with \boldmath{\texorpdfstring{$k_{\rm h}$}{kh} or \texorpdfstring{$k_{\rm S}$}{kS}}}

In the $k_{\rm S}$-damping model, the function $\Theta(t,k)$ is given by
\begin{equation}
    \Theta(t, k) = \theta(k_{\rm S}(t) - k) \theta( k_{\rm S}(t) - a(t) H(t))\, .
    \label{eq: kS-model}
\end{equation}
Here, the wavenumber $k_{\rm S}(t)$ characterizes the typical particle--antiparticle separation following their creation. 
Hence, the first factor in Eq.~\eqref{eq: kS-model} ensures that gauge-field modes with $k > k_{\rm S}$ are not affected by the conductive medium,
while the second factor prohibits super-Hubble particle creation.

The scale $k_{\rm S}$ can be determined from a kinematic analysis. For \FAISM, one has~\cite{eckardstein_2025}
\begin{equation}
    k_{\rm S} = a\frac{C^{1/3}g'(\mu)}{2^{1/4}} \left[\langle{\bm E}^2\rangle -\langle{\bm B}^2\rangle + \sqrt{\langle{\bm E}^2 -{\bm B}^2\rangle^{^2} + 4\langle \bm{E}\cdot \bm{B}\rangle^{^2}}\right]^{1/4}\, .
    \label{eq: kS}
\end{equation}
However, as explained in Ref.~\cite{eckardstein_2025}, the $k_{\rm S}$-damping model is more challenging to implement in the GEF. In particular, in the present analysis, we found it to be unsuitable for a parameter scan like the one in Sec.~\ref{sec: Results}. The challenge arises as there is a second important scale affecting gauge-field evolution besides the damping scale $k_{\rm S}$: the instability scale $k_{\rm h}$, which determines if a gauge-field mode $A_\lambda(t,k)$ is tachyonically amplified. It is given by~\cite{eckardstein_2025}
\begin{equation}
    k_{\rm h}(t) = \underset{t' < t}{\operatorname{max}} \left\{ a(t')H(t') \left(|\xi_{\rm eff}| + \sqrt{\xi_{\rm eff}^2 + s_{E}^2+ s_{E}} \right) \right\} \,, \qquad s_{E} = \frac{\sigma_{E}\Theta}{2H} \,.
    \label{eq: kh}
\end{equation}
Since $k_{\rm h}$ and $k_{\rm S}$ are evidently different, when modeling $\Theta(t,k)$ as in Eq.~\eqref{eq: kS-model}, one consistently needs to distinguish between gauge-field modes that have already experienced the tachyonic instability before feeling the presence of the damping conductive medium, and those modes that are damped before being tachyonically amplified. 
Therefore, the GEF modeling becomes significantly more involved, as we have explained in detail in Ref.~\cite{eckardstein_2025}.

To facilitate the numerical implementation of $\Theta(t,k)$, one can make the approximate identification $k_{\rm h} \simeq k_{\rm S}$, such that%
\footnote{Note that only in the first Heaviside function $k_{\rm h}$ is identified with $k_{\rm S}$. The second, $k$-independent Heaviside function affects all gauge modes $A_\lambda(t,k)$ equally. Thus, we do not gain anything in terms of the numerical implementation by approximating $k_{\rm S} \simeq k_{\rm h}$ for this second term.}
\begin{equation}
    \Theta(t, k) = \theta(k_{\rm h}(t) - k) \theta( k_{\rm S}(t) - a(t) H(t))\, .
    \label{eq: kh-model}
\end{equation}
This second choice is what we refer to as the $k_{\rm h}$-damping model.

Just comparing Eqs.~\eqref{eq: kS} and \eqref{eq: kh}, it may not be evident why this identification is justified. One approach is to verify the assumption \textit{a posteriori}. However, given that both quantities scale with $a H$, it is not surprising that one will  find them to differ only by an $\mathcal{O}(1)$ factor, which is insignificant when considering the exponential growth of the scale factor during inflation.

\subsection[Comparison between \texorpdfstring{$k_{\rm h}$}{kh}- and \texorpdfstring{$k_{\rm S}$}{kS}-damping]{Comparison between \boldmath{\texorpdfstring{$k_{\rm h}$}{kh}- and \texorpdfstring{$k_{\rm S}$}{kS}}-damping}

In order to justify that we rely on the simpler $k_{\rm h}$-damping model in the main body of the text, we now study how our main results are affected by this choice.

\begin{figure}
    \centering
    \includegraphics[width=0.98\textwidth]{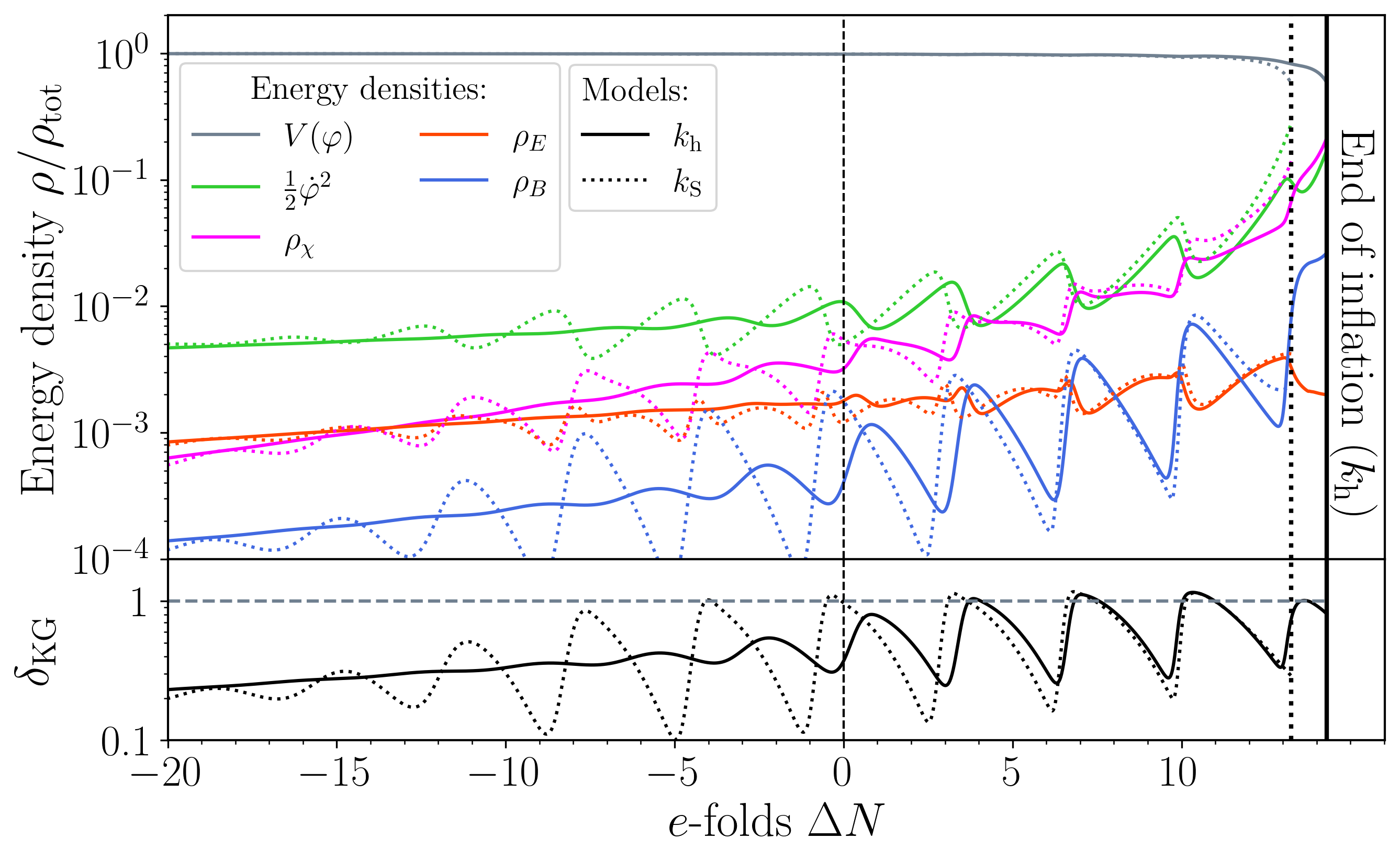}
    \caption
    {
       Same as in Fig.~\ref{fig: Schwinger BR}, 
       but we overlay the previous results for $k_{\rm h}$-damping (solid lines) with those for $k_{\rm S}$-damping (dotted lines).
     }
    \label{fig: Schwinger BR - kS}
\end{figure}

First, we re-analyze the onset of the regime of fermion-tempered backreaction by comparing the results for the benchmark point $\beta = 59$ and  $m= \num{2.5e-5} \MP$ as given in Fig.~\ref{fig: Schwinger BR}.
The updated version of this plot can be seen in Fig.~\ref{fig: Schwinger BR - kS}, where we overlay the previous results for $k_{\rm h}$-damping (solid lines) with those for $k_{\rm S}$-damping (dotted lines).
Notably, the backreaction qualitatively remains the same: an oscillation of the inflaton's kinetic energy around the slow-roll attractor accompanied by oscillations in $\rho_\chi$ and $\rho_{\rm EM}$.
However, we do see that backreaction sets in marginally earlier in the $k_{\rm S}$-damping model. We can understand this by considering that $k_{\rm S} < k_{\rm h}$ at early times, and, hence, some gauge modes $A_\lambda(t,k)$ will be tachyonically amplified without immediately feeling the damping effect due to the fermion fluid. This implies that the integrated gauge field can start to affect the inflaton dynamics earlier.
At the same time, the end of inflation occurs marginally earlier  in the  $k_{\rm S}$-damping model (the vertical dotted line at $\Delta N \simeq 13$). We may interpret this by considering that, once $k_{\rm S} > k_{\rm h}$, some gauge modes will be damped even before feeling the enhancing effect of the rolling inflaton field. 
Overall, we are reassured in the fact that fermion-tempered backreaction is not qualitatively affected by our modeling of $\Theta(t,k)$.

Next, we show how the SGWB spectra in Fig.~\ref{fig: spectra, inst. reh} are affected by our modeling of $\Theta(t,k)$. We show the updated version of this plot in Fig.~\ref{fig: spectra, inst. reh - kS}. Again, we overlay the original result
for $k_{\rm h}$-damping (dashed lines) with those for $k_{\rm S}$-damping (dotted lines).
Again, we find good qualitative agreement for the GW spectra corresponding to the benchmark points $\beta=59$ and $m=\num{1.5e-6} \MP$ (cyan), $\num{2.5e-6} \MP$ (violet). In both models, they reach roughly the same amplitude, cover the same frequency range, are detectable by the same observatories, and feature the same spectral slope. For the $k_{\rm S}$-damping model, the amplitudes of the spectra are marginally decreased, and the growth is marginally less stable, including some small oscillations.
For $\beta = 59$ and $m= \num{2.5e-5} \MP$ we also find qualitative agreement between the results for $k_{\rm h}$-damping and $k_{\rm S}$-damping. Both spectra reach NG15 sensitivity, but are ruled out by PLANCK. 
However, as one can anticipate based on our results in Fig.~\ref{fig: Schwinger BR - kS}, the spectra for $k_{\rm S}$-damping feature strong oscillations already at lower frequencies. Also, one can see the slight reduction of redshift 
due to $\Delta N_{\rm BR}$ from the later rise in the GFIGW component at low frequencies for the $k_{\rm S}$-damping model. 
Again, the results are in overall good qualitative agreement, and we find that the results are consistent with the interpretation given in the main body of the text.
\begin{figure}
    \centering
    \includegraphics[width=0.95\textwidth]{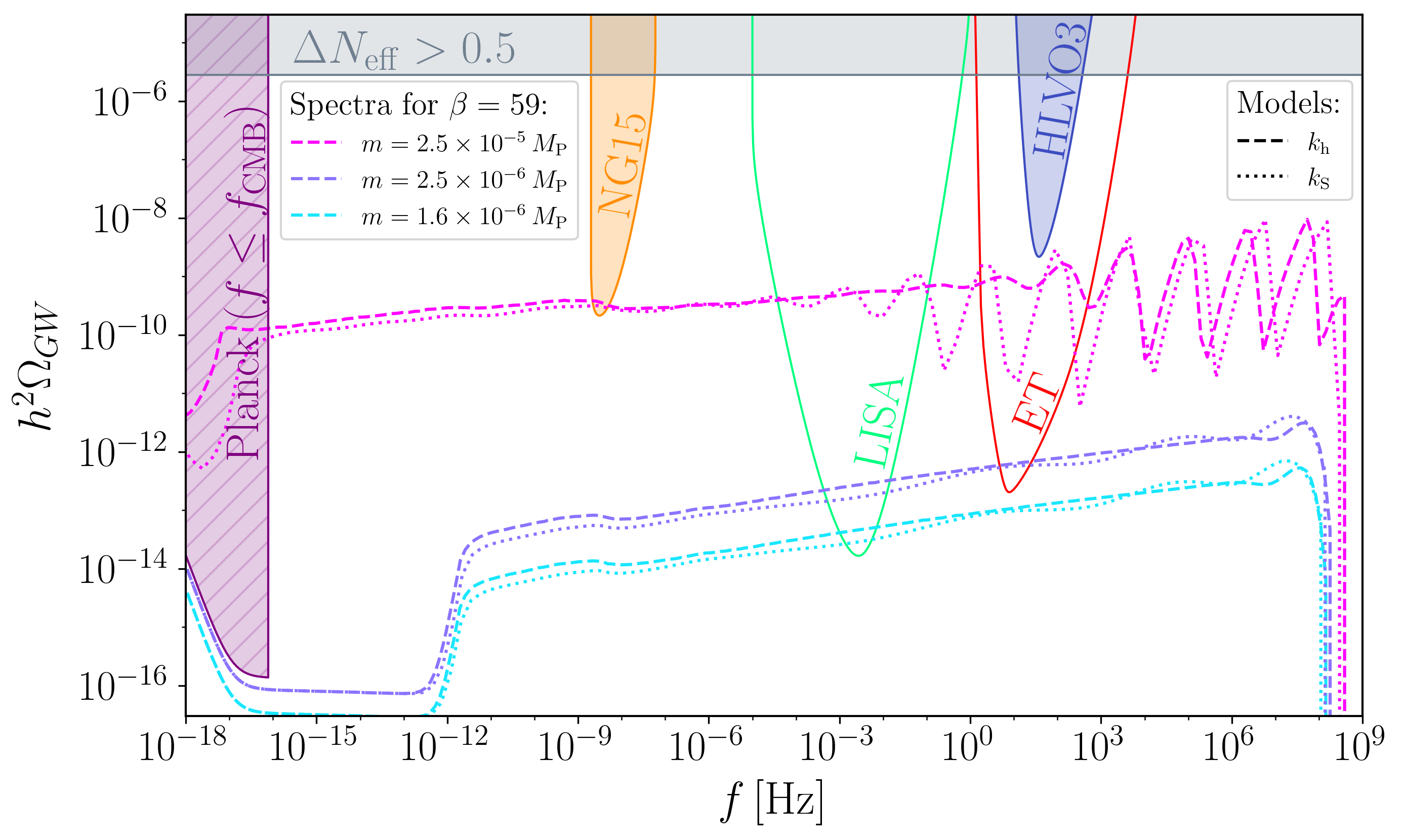}
    \caption
    {
       Same as in Fig.~\ref{fig: spectra, inst. reh}, 
       but we overlay the previous results for \textit{$k_{\rm h}$-damping} (dashed lines) with those for \textit{$k_{\rm S}$-damping} (dotted lines).
     }
    \label{fig: spectra, inst. reh - kS}
\end{figure}

The fact that the two models for $\Theta(t,k)$ give such similar results can also be understood quantitatively. We find that, once $k_{\rm S} > k_{\rm h}$, they only ever differ at most by a factor of two, while both scales vary exponentially with time.
Therefore, it does not matter much if one distinguishes between $k_{\rm S}$ or $k_{\rm h}$ when modeling the scale dependence of the Schwinger induced current in Eq.~\eqref{eq: ScaleDep Ohmic current}.
The agreement between the two models is especially satisfactory given that we treat the fermions effectively instead of relying on a first-principles computation. These encouraging observations notwithstanding, we also argue that it would be interesting to investigate the new regime of fermion-tempered backreaction on the lattice in order to resolve the remaining conceptual uncertainties inherent to our approach.


\bibliographystyle{JHEP}
\bibliography{manuscript_2.bib}

@article{Sauter_1931,
    author = "Sauter, Fritz",
    title = "{\"Uber das Verhalten eines Elektrons im homogenen elektrischen Feld nach der relativistischen Theorie Diracs}",
    doi = "10.1007/BF01339461",
    journal = "Z. Phys.",
    volume = "69",
    pages = "742--764",
    year = "1931"
}

@article{Heisenberg_1936,
    author = "Heisenberg, W. and Euler, H.",
    title = "{Folgerungen aus der Diracschen Theorie des Positrons}",
    eprint = "physics/0605038",
    archivePrefix = "arXiv",
    doi = "10.1007/BF01343663",
    journal = "Z. Phys.",
    volume = "98",
    number = "11-12",
    pages = "714--732",
    year = "1936"
}

@article{Schwinger_1951,
    author = "Schwinger, Julian S.",
    editor = "Milton, K. A.",
    title = "{On gauge invariance and vacuum polarization}",
    doi = "10.1103/PhysRev.82.664",
    journal = "Phys. Rev.",
    volume = "82",
    pages = "664--679",
    year = "1951"
}

@article{starobinsky_1980,
    author = "Starobinsky, Alexei A.",
    editor = "Khalatnikov, I. M. and Mineev, V. P.",
    title = "{A New Type of Isotropic Cosmological Models Without Singularity}",
    doi = "10.1016/0370-2693(80)90670-X",
    journal = "Phys. Lett. B",
    volume = "91",
    pages = "99--102",
    year = "1980"
}

@article{guth_1981,
    author = "Guth, Alan H.",
    editor = "Fang, Li-Zhi and Ruffini, R.",
    title = "{The Inflationary Universe: A Possible Solution to the Horizon and Flatness Problems}",
    reportNumber = "SLAC-PUB-2576",
    doi = "10.1103/PhysRevD.23.347",
    journal = "Phys. Rev. D",
    volume = "23",
    pages = "347--356",
    year = "1981"
}

@article{Mukhanov_1981,
    author = "Mukhanov, Viatcheslav F. and Chibisov, G. V.",
    title = "{Quantum Fluctuations and a Nonsingular Universe}",
    journal = "JETP Lett.",
    volume = "33",
    pages = "532--535",
    year = "1981"
}

@article{linde_1982,
    author = "Linde, Andrei D.",
    editor = "Fang, Li-Zhi and Ruffini, R.",
    title = "{A New Inflationary Universe Scenario: A Possible Solution of the Horizon, Flatness, Homogeneity, Isotropy and Primordial Monopole Problems}",
    reportNumber = "LEBEDEV-81-229",
    doi = "10.1016/0370-2693(82)91219-9",
    journal = "Phys. Lett. B",
    volume = "108",
    pages = "389--393",
    year = "1982"
}

@article{Mukhanov_1982,
    author = "Mukhanov, Viatcheslav F. and Chibisov, G. V.",
    title = "{The Vacuum energy and large scale structure of the universe}",
    journal = "Sov. Phys. JETP",
    volume = "56",
    pages = "258--265",
    year = "1982"
}

@article{albrecht_1982,
    author = "Albrecht, Andreas and Steinhardt, Paul J.",
    editor = "Fang, Li-Zhi and Ruffini, R.",
    title = "{Cosmology for Grand Unified Theories with Radiatively Induced Symmetry Breaking}",
    reportNumber = "UPR-0185T",
    doi = "10.1103/PhysRevLett.48.1220",
    journal = "Phys. Rev. Lett.",
    volume = "48",
    pages = "1220--1223",
    year = "1982"
}

@article{hawking_1982,
    author = "Hawking, S. W.",
    title = "{The Development of Irregularities in a Single Bubble Inflationary Universe}",
    reportNumber = "Print-83-0015 (CAMBRIDGE)",
    doi = "10.1016/0370-2693(82)90373-2",
    journal = "Phys. Lett. B",
    volume = "115",
    pages = "295",
    year = "1982"
}

@article{guth_1982,
    author = "Guth, Alan H. and Pi, S. Y.",
    title = "{Fluctuations in the New Inflationary Universe}",
    doi = "10.1103/PhysRevLett.49.1110",
    journal = "Phys. Rev. Lett.",
    volume = "49",
    pages = "1110--1113",
    year = "1982"
}

@article{starobinsky_1982,
    author = "Starobinsky, Alexei A.",
    title = "{Dynamics of Phase Transition in the New Inflationary Universe Scenario and Generation of Perturbations}",
    doi = "10.1016/0370-2693(82)90541-X",
    journal = "Phys. Lett. B",
    volume = "117",
    pages = "175--178",
    year = "1982"
}

@article{bardeen_1983,
    author = "Bardeen, James M. and Steinhardt, Paul J. and Turner, Michael S.",
    title = "{Spontaneous Creation of Almost Scale - Free Density Perturbations in an Inflationary Universe}",
    reportNumber = "UPR-0202T, EFI-83-13-CHICAGO",
    doi = "10.1103/PhysRevD.28.679",
    journal = "Phys. Rev. D",
    volume = "28",
    pages = "679",
    year = "1983"
}

@article{linde_1983,
    author = "Linde, Andrei D.",
    title = "{Chaotic Inflation}",
    doi = "10.1016/0370-2693(83)90837-7",
    journal = "Phys. Lett. B",
    volume = "129",
    pages = "177--181",
    year = "1983"
}

@article{freese_1990,
    author = "Freese, Katherine and Frieman, Joshua A. and Olinto, Angela V.",
    title = "{Natural inflation with pseudo - Nambu-Goldstone bosons}",
    reportNumber = "FERMILAB-PUB-90-177-A",
    doi = "10.1103/PhysRevLett.65.3233",
    journal = "Phys. Rev. Lett.",
    volume = "65",
    pages = "3233--3236",
    year = "1990"
}

@article{garretson_1992,
    author = "Garretson, W. Daniel and Field, George B. and Carroll, Sean M.",
    title = "{Primordial magnetic fields from pseudoGoldstone bosons}",
    eprint = "hep-ph/9209238",
    archivePrefix = "arXiv",
    reportNumber = "PRINT-92-0448 (CFA,CAMBRIDGE), CFA-3507",
    doi = "10.1103/PhysRevD.46.5346",
    journal = "Phys. Rev. D",
    volume = "46",
    pages = "5346--5351",
    year = "1992"
}

@inproceedings{allen_1996,
    author = "Allen, Bruce",
    title = "{The Stochastic gravity wave background: Sources and detection}",
    booktitle = "{Les Houches School of Physics: Astrophysical Sources of Gravitational Radiation}",
    eprint = "gr-qc/9604033",
    archivePrefix = "arXiv",
    reportNumber = "WISC-MILW-96-TH-22",
    pages = "373--417",
    month = "4",
    year = "1996"
}

@article{allen_1997,
    author = "Allen, Bruce and Romano, Joseph D.",
    title = "{Detecting a stochastic background of gravitational radiation: Signal processing strategies and sensitivities}",
    eprint = "gr-qc/9710117",
    archivePrefix = "arXiv",
    reportNumber = "WISC-MILW-97-TH-14",
    doi = "10.1103/PhysRevD.59.102001",
    journal = "Phys. Rev. D",
    volume = "59",
    pages = "102001",
    year = "1999"
}

@article{maggiore_2000,
    author = "Maggiore, Michele",
    title = "{Gravitational wave experiments and early universe cosmology}",
    eprint = "gr-qc/9909001",
    archivePrefix = "arXiv",
    reportNumber = "IFUP-TH-20-99",
    doi = "10.1016/S0370-1573(99)00102-7",
    journal = "Phys. Rept.",
    volume = "331",
    pages = "283--367",
    year = "2000"
}

@article{anber_2006,
    author = "Anber, Mohamed M. and Sorbo, Lorenzo",
    title = "{N-flationary magnetic fields}",
    eprint = "astro-ph/0606534",
    archivePrefix = "arXiv",
    doi = "10.1088/1475-7516/2006/10/018",
    journal = "JCAP",
    volume = "10",
    pages = "018",
    year = "2006"
}

@article{Anber_2010,
   title={Naturally inflating on steep potentials through electromagnetic dissipation},
   volume={81},
   ISSN={1550-2368},
   url={http://dx.doi.org/10.1103/PhysRevD.81.043534},
   DOI={10.1103/physrevd.81.043534},
   number={4},
   journal={Physical Review D},
   publisher={American Physical Society (APS)},
   author={Anber, Mohamed M. and Sorbo, Lorenzo},
   year={2010},
   month=feb 
}

@article{LIGO_2010,
	 author = "Harry, Gregory M.",
    editor = "Marka, Zsuzsa and Marka, Szabolcs",
    collaboration = "LIGO Scientific",
    title = "{Advanced LIGO: The next generation of gravitational wave detectors}",
    doi = "10.1088/0264-9381/27/8/084006",
    journal = "Class. Quant. Grav.",
    volume = "27",
    pages = "084006",
    year = "2010"
}

@article{ET_2010,
	author = "Punturo, M. and others",
    editor = "Ricci, Fulvio",
    title = "{The Einstein Telescope: A third-generation gravitational wave observatory}",
    doi = "10.1088/0264-9381/27/19/194002",
    journal = "Class. Quant. Grav.",
    volume = "27",
    pages = "194002",
    year = "2010"
}

@article{Barnaby_2011_A,
    author = "Barnaby, Neil and Peloso, Marco",
    title = "{Large Nongaussianity in Axion Inflation}",
    eprint = "1011.1500",
    archivePrefix = "arXiv",
    primaryClass = "hep-ph",
    reportNumber = "UMN-TH-2926-10",
    doi = "10.1103/PhysRevLett.106.181301",
    journal = "Phys. Rev. Lett.",
    volume = "106",
    pages = "181301",
    year = "2011"
}

@article{Barnaby_2011_B,
    author = "Barnaby, Neil and Namba, Ryo and Peloso, Marco",
    title = "{Phenomenology of a Pseudo-Scalar Inflaton: Naturally Large Nongaussianity}",
    eprint = "1102.4333",
    archivePrefix = "arXiv",
    primaryClass = "astro-ph.CO",
    doi = "10.1088/1475-7516/2011/04/009",
    journal = "JCAP",
    volume = "04",
    pages = "009",
    year = "2011"
}

@article{Adshead_2012,
    author = "Adshead, Peter and Wyman, Mark",
    title = "{Chromo-Natural Inflation: Natural inflation on a steep potential with classical non-Abelian gauge fields}",
    eprint = "1202.2366",
    archivePrefix = "arXiv",
    primaryClass = "hep-th",
    doi = "10.1103/PhysRevLett.108.261302",
    journal = "Phys. Rev. Lett.",
    volume = "108",
    pages = "261302",
    year = "2012"
}

@article{Frob_2014,
    author = {Fr{\"o}b, Markus B. and Garriga, Jaume and Kanno, Sugumi and Sasaki, Misao and Soda, Jiro and Tanaka, Takahiro and Vilenkin, Alexander},
    title = "{Schwinger effect in de Sitter space}",
    eprint = "1401.4137",
    archivePrefix = "arXiv",
    primaryClass = "hep-th",
    reportNumber = "QGASLAB-14-01, KOBE-TH-14-01",
    doi = "10.1088/1475-7516/2014/04/009",
    journal = "JCAP",
    volume = "04",
    pages = "009",
    year = "2014"
}

@article{Kobayashi_2014,
    author = "Kobayashi, Takeshi and Afshordi, Niayesh",
    title = "{Schwinger Effect in 4D de Sitter Space and Constraints on Magnetogenesis in the Early Universe}",
    eprint = "1408.4141",
    archivePrefix = "arXiv",
    primaryClass = "hep-th",
    doi = "10.1007/JHEP10(2014)166",
    journal = "JHEP",
    volume = "10",
    pages = "166",
    year = "2014"
}

@article{VIRGO_2014,
	author = "Acernese, F. and others",
    collaboration = "VIRGO",
    title = "{Advanced Virgo: a second-generation interferometric gravitational wave detector}",
    eprint = "1408.3978",
    archivePrefix = "arXiv",
    primaryClass = "gr-qc",
    doi = "10.1088/0264-9381/32/2/024001",
    journal = "Class. Quant. Grav.",
    volume = "32",
    number = "2",
    pages = "024001",
    year = "2015"
}

@article{LIGO_2014,
	 author = "Aasi, J. and others",
    collaboration = "LIGO Scientific",
    title = "{Advanced LIGO}",
    eprint = "1411.4547",
    archivePrefix = "arXiv",
    primaryClass = "gr-qc",
    doi = "10.1088/0264-9381/32/7/074001",
    journal = "Class. Quant. Grav.",
    volume = "32",
    pages = "074001",
    year = "2015"
}

@article{Kuroyanagi_2014,
    author = "Kuroyanagi, Sachiko and Takahashi, Tomo and Yokoyama, Shuichiro",
    title = "{Blue-tilted Tensor Spectrum and Thermal History of the Universe}",
    eprint = "1407.4785",
    archivePrefix = "arXiv",
    primaryClass = "astro-ph.CO",
    reportNumber = "ICRR-REPORT-686-2014-12",
    doi = "10.1088/1475-7516/2015/02/003",
    journal = "JCAP",
    volume = "02",
    pages = "003",
    year = "2015"
}

@article{Stahl_2015,
    author = "Stahl, Cl{\'e}ment and Strobel, Eckhard and Xue, She-Sheng",
    title = "{Fermionic current and Schwinger effect in de Sitter spacetime}",
    eprint = "1507.01686",
    archivePrefix = "arXiv",
    primaryClass = "gr-qc",
    doi = "10.1103/PhysRevD.93.025004",
    journal = "Phys. Rev. D",
    volume = "93",
    number = "2",
    pages = "025004",
    year = "2016"
}

@article{Hayashinaka_2016_A,
    author = "Hayashinaka, Takahiro and Yokoyama, Jun'ichi",
    title = "{Point splitting renormalization of Schwinger induced current in de Sitter spacetime}",
    eprint = "1603.06172",
    archivePrefix = "arXiv",
    primaryClass = "hep-th",
    reportNumber = "RESCEU-13-16",
    doi = "10.1088/1475-7516/2016/07/012",
    journal = "JCAP",
    volume = "07",
    pages = "012",
    year = "2016"
}

@article{Hayashinaka_2016_B,
    author = "Hayashinaka, Takahiro and Fujita, Tomohiro and Yokoyama, Jun'ichi",
    title = "{Fermionic Schwinger effect and induced current in de Sitter space}",
    eprint = "1603.04165",
    archivePrefix = "arXiv",
    primaryClass = "hep-th",
    reportNumber = "RESCEU-12-16",
    doi = "10.1088/1475-7516/2016/07/010",
    journal = "JCAP",
    volume = "07",
    pages = "010",
    year = "2016"
}

@article{Bavarsad_2016,
    author = "Bavarsad, Ehsan and Stahl, Cl{\'e}ment and Xue, She-Sheng",
    title = "{Scalar current of created pairs by Schwinger mechanism in de Sitter spacetime}",
    eprint = "1602.06556",
    archivePrefix = "arXiv",
    primaryClass = "hep-th",
    doi = "10.1103/PhysRevD.94.104011",
    journal = "Phys. Rev. D",
    volume = "94",
    number = "10",
    pages = "104011",
    year = "2016"
}

@misc{LISA_2017,
	 author = "Amaro-Seoane, Pau and others",
    collaboration = "LISA",
    title = "{Laser Interferometer Space Antenna}",
    eprint = "1702.00786",
    archivePrefix = "arXiv",
    primaryClass = "astro-ph.IM",
    month = "2",
    year = "2017"
}

@article{Banyeres_2018,
    author = "Banyeres, Mariona and Dom{\`e}nech, Guillem and Garriga, Jaume",
    title = "{Vacuum birefringence and the Schwinger effect in (3+1) de Sitter}",
    eprint = "1809.08977",
    archivePrefix = "arXiv",
    primaryClass = "hep-th",
    doi = "10.1088/1475-7516/2018/10/023",
    journal = "JCAP",
    volume = "10",
    pages = "023",
    year = "2018"
}

@article{caprini_2018,
    author = "Caprini, Chiara and Figueroa, Daniel G.",
    title = "{Cosmological Backgrounds of Gravitational Waves}",
    eprint = "1801.04268",
    archivePrefix = "arXiv",
    primaryClass = "astro-ph.CO",
    doi = "10.1088/1361-6382/aac608",
    journal = "Class. Quant. Grav.",
    volume = "35",
    number = "16",
    pages = "163001",
    year = "2018"
}

@article{domcke_2018,
    author = "Domcke, Valerie and Mukaida, Kyohei",
    title = "{Gauge Field and Fermion Production during Axion Inflation}",
    eprint = "1806.08769",
    archivePrefix = "arXiv",
    primaryClass = "hep-ph",
    reportNumber = "DESY 18-098, DESY-18-098",
    doi = "10.1088/1475-7516/2018/11/020",
    journal = "JCAP",
    volume = "11",
    pages = "020",
    year = "2018"
}

@article{Hayashinaka_2018,
    author = "Hayashinaka, Takahiro and Xue, She-Sheng",
    title = "{Physical renormalization condition for de Sitter QED}",
    eprint = "1802.03686",
    archivePrefix = "arXiv",
    primaryClass = "gr-qc",
    reportNumber = "RESCEU-3-18",
    doi = "10.1103/PhysRevD.97.105010",
    journal = "Phys. Rev. D",
    volume = "97",
    number = "10",
    pages = "105010",
    year = "2018"
}

@article{Sobol_2018,
    author = "Sobol, O. O. and Gorbar, E. V. and Kamarpour, M. and Vilchinskii, S. I.",
    title = "{Influence of backreaction of electric fields and Schwinger effect on inflationary magnetogenesis}",
    eprint = "1807.09851",
    archivePrefix = "arXiv",
    primaryClass = "hep-ph",
    doi = "10.1103/PhysRevD.98.063534",
    journal = "Phys. Rev. D",
    volume = "98",
    number = "6",
    pages = "063534",
    year = "2018"
}

@article{Kitamoto_2018,
    author = "Kitamoto, Hiroyuki",
    title = "{Schwinger Effect in Inflaton-Driven Electric Field}",
    eprint = "1807.03753",
    archivePrefix = "arXiv",
    primaryClass = "hep-th",
    reportNumber = "NCTS-TH/1809, NCTS-TH-1809",
    doi = "10.1103/PhysRevD.98.103512",
    journal = "Phys. Rev. D",
    volume = "98",
    number = "10",
    pages = "103512",
    year = "2018"
}

@article{Gorbar_2019,
    author = "Gorbar, E. V. and Momot, A. I. and Sobol, O. O. and Vilchinskii, S. I.",
    title = "{Kinetic approach to the Schwinger effect during inflation}",
    eprint = "1909.10332",
    archivePrefix = "arXiv",
    primaryClass = "gr-qc",
    doi = "10.1103/PhysRevD.100.123502",
    journal = "Phys. Rev. D",
    volume = "100",
    number = "12",
    pages = "123502",
    year = "2019"
}

@misc{LISA_2019,
	author = "Baker, John and others",
    title = "{The Laser Interferometer Space Antenna: Unveiling the Millihertz Gravitational Wave Sky}",
    eprint = "1907.06482",
    archivePrefix = "arXiv",
    primaryClass = "astro-ph.IM",
    reportNumber = "FERMILAB-PUB-19-436-A",
    month = "7",
    year = "2019"
}

@article{Sobol_2019,
    author = "Sobol, O. O. and Gorbar, E. V. and Vilchinskii, S. I.",
    title = "{Backreaction of electromagnetic fields and the Schwinger effect in pseudoscalar inflation magnetogenesis}",
    eprint = "1907.10443",
    archivePrefix = "arXiv",
    primaryClass = "astro-ph.CO",
    doi = "10.1103/PhysRevD.100.063523",
    journal = "Phys. Rev. D",
    volume = "100",
    number = "6",
    pages = "063523",
    year = "2019"
}

@article{planck_2020_VI,
	author = "Aghanim, N. and others",
    collaboration = "Planck",
    title = "{Planck 2018 results. VI. Cosmological parameters}",
    eprint = "1807.06209",
    archivePrefix = "arXiv",
    primaryClass = "astro-ph.CO",
    doi = "10.1051/0004-6361/201833910",
    journal = "Astron. Astrophys.",
    volume = "641",
    pages = "A6",
    year = "2020",
    note = "[Erratum: Astron.Astrophys. 652, C4 (2021)]"
}

@article{planck_2020_X,
    author = "Akrami, Y. and others",
    collaboration = "Planck",
    title = "{Planck 2018 results. X. Constraints on inflation}",
    eprint = "1807.06211",
    archivePrefix = "arXiv",
    primaryClass = "astro-ph.CO",
    doi = "10.1051/0004-6361/201833887",
    journal = "Astron. Astrophys.",
    volume = "641",
    pages = "A10",
    year = "2020"
}

@article{Kuroyanagi_2020,
    author = "Kuroyanagi, Sachiko and Takahashi, Tomo and Yokoyama, Shuichiro",
    title = "{Blue-tilted inflationary tensor spectrum and reheating in the light of NANOGrav results}",
    eprint = "2011.03323",
    archivePrefix = "arXiv",
    primaryClass = "astro-ph.CO",
    doi = "10.1088/1475-7516/2021/01/071",
    journal = "JCAP",
    volume = "01",
    pages = "071",
    year = "2021"
}

@article{Domcke_2020_Resonant,
    author = "Domcke, Valerie and Guidetti, Veronica and Welling, Yvette and Westphal, Alexander",
    title = "{Resonant backreaction in axion inflation}",
    eprint = "2002.02952",
    archivePrefix = "arXiv",
    primaryClass = "astro-ph.CO",
    reportNumber = "DESY-20-017",
    doi = "10.1088/1475-7516/2020/09/009",
    journal = "JCAP",
    volume = "09",
    pages = "009",
    year = "2020"
}

@article{Domcke_2020_Fermions,
    author = "Domcke, Valerie and Ema, Yohei and Mukaida, Kyohei",
    title = "{Chiral Anomaly, Schwinger Effect, Euler-Heisenberg Lagrangian, and application to axion inflation}",
    eprint = "1910.01205",
    archivePrefix = "arXiv",
    primaryClass = "hep-ph",
    reportNumber = "DESY-19-166, DESY 19-166",
    doi = "10.1007/JHEP02(2020)055",
    journal = "JHEP",
    volume = "02",
    pages = "055",
    year = "2020"
}

@article{Sobol_2020,
    author = "Sobol, O. O. and Gorbar, E. V. and Momot, A. I. and Vilchinskii, S. I.",
    title = "{Schwinger production of scalar particles during and after inflation from the first principles}",
    eprint = "2004.12664",
    archivePrefix = "arXiv",
    primaryClass = "gr-qc",
    doi = "10.1103/PhysRevD.102.023506",
    journal = "Phys. Rev. D",
    volume = "102",
    number = "2",
    pages = "023506",
    year = "2020"
}

@article{schmitz_2021,
    author = "Schmitz, Kai",
    title = "{New Sensitivity Curves for Gravitational-Wave Signals from Cosmological Phase Transitions}",
    eprint = "2002.04615",
    archivePrefix = "arXiv",
    primaryClass = "hep-ph",
    reportNumber = "CERN-TH-2020-018",
    doi = "10.1007/JHEP01(2021)097",
    journal = "JHEP",
    volume = "01",
    pages = "097",
    year = "2021"
}

@article{yeh_2021,
    author = "Yeh, Tsung-Han and Olive, Keith A. and Fields, Brian D.",
    title = "{The impact of new $d(p,\gamma)$3 rates on Big Bang Nucleosynthesis}",
    eprint = "2011.13874",
    archivePrefix = "arXiv",
    primaryClass = "astro-ph.CO",
    reportNumber = "UMN--TH--4004/20, FTPI--MINN--20/35",
    doi = "10.1088/1475-7516/2021/03/046",
    journal = "JCAP",
    volume = "03",
    pages = "046",
    year = "2021"
}

@article{pisanti_2021,
    author = "Pisanti, Ofelia and Mangano, Gianpiero and Miele, Gennaro and Mazzella, Pierpaolo",
    title = "{Primordial Deuterium after LUNA: concordances and error budget}",
    eprint = "2011.11537",
    archivePrefix = "arXiv",
    primaryClass = "astro-ph.CO",
    doi = "10.1088/1475-7516/2021/04/020",
    journal = "JCAP",
    volume = "04",
    pages = "020",
    year = "2021"
}

@article{LIGO_collaboration_2021,
	 author = "Abbott, R. and others",
    collaboration = "KAGRA, Virgo, LIGO Scientific",
    title = "{Upper limits on the isotropic gravitational-wave background from Advanced LIGO and Advanced Virgo{\textquoteright}s third observing run}",
    eprint = "2101.12130",
    archivePrefix = "arXiv",
    primaryClass = "gr-qc",
    reportNumber = "LIGO-DCC-P2000314",
    doi = "10.1103/PhysRevD.104.022004",
    journal = "Phys. Rev. D",
    volume = "104",
    number = "2",
    pages = "022004",
    year = "2021"
}

@article{Gorbar_2021,
    author = "Gorbar, E. V. and Schmitz, K. and Sobol, O. O. and Vilchinskii, S. I.",
    title = "{Gauge-field production during axion inflation in the gradient expansion formalism}",
    eprint = "2109.01651",
    archivePrefix = "arXiv",
    primaryClass = "hep-ph",
    reportNumber = "CERN-TH-2021-128",
    doi = "10.1103/PhysRevD.104.123504",
    journal = "Phys. Rev. D",
    volume = "104",
    number = "12",
    pages = "123504",
    year = "2021"
}

@article{gorbar_2022,
    author = "Gorbar, E. V. and Schmitz, K. and Sobol, O. O. and Vilchinskii, S. I.",
    title = "{Hypermagnetogenesis from axion inflation: Model-independent estimates}",
    eprint = "2111.04712",
    archivePrefix = "arXiv",
    primaryClass = "hep-ph",
    reportNumber = "CERN-TH-2021-185",
    doi = "10.1103/PhysRevD.105.043530",
    journal = "Phys. Rev. D",
    volume = "105",
    number = "4",
    pages = "043530",
    year = "2022"
}

@article{tristram_2022,
	author = "Tristram, M. and others",
    title = "{Improved limits on the tensor-to-scalar ratio using BICEP and Planck data}",
    eprint = "2112.07961",
    archivePrefix = "arXiv",
    primaryClass = "astro-ph.CO",
    doi = "10.1103/PhysRevD.105.083524",
    journal = "Phys. Rev. D",
    volume = "105",
    number = "8",
    pages = "083524",
    year = "2022"
}

@article{fujita_2022,
    author = "Fujita, Tomohiro and Kume, Jun'ya and Mukaida, Kyohei and Tada, Yuichiro",
    title = "{Effective treatment of U(1) gauge field and charged particles in axion inflation}",
    eprint = "2204.01180",
    archivePrefix = "arXiv",
    primaryClass = "hep-ph",
    reportNumber = "RESCEU-3/22, KEK-TH-2402",
    doi = "10.1088/1475-7516/2022/09/023",
    journal = "JCAP",
    volume = "09",
    pages = "023",
    year = "2022"
}

@article{cado_2022,
    author = "Cado, Yann and Quir{\'o}s, Mariano",
    title = "{Numerical study of the Schwinger effect in axion inflation}",
    eprint = "2208.10977",
    archivePrefix = "arXiv",
    primaryClass = "hep-ph",
    doi = "10.1103/PhysRevD.106.123527",
    journal = "Phys. Rev. D",
    volume = "106",
    number = "12",
    pages = "123527",
    year = "2022"
}

@article{domcke_2023,
    author = "Domcke, Valerie and Kamada, Kohei and Mukaida, Kyohei and Schmitz, Kai and Yamada, Masaki",
    title = "{Wash-in leptogenesis after axion inflation}",
    eprint = "2210.06412",
    archivePrefix = "arXiv",
    primaryClass = "hep-ph",
    reportNumber = "CERN-TH-2022-162, RESCEU-17/22, KEK-TH-2455, MS-TP-22-37, TU-1170",
    doi = "10.1007/JHEP01(2023)053",
    journal = "JHEP",
    volume = "01",
    pages = "053",
    year = "2023"
}

@article{gorbar_2023,
    author = "Gorbar, E. V. and Momot, A. I. and Rudenok, I. V. and Sobol, O. O. and Vilchinskii, S. I. and Oleinikova, I. V.",
    title = "{Chirality Production during Axion Inflation}",
    eprint = "2111.05848",
    archivePrefix = "arXiv",
    primaryClass = "hep-ph",
    doi = "10.15407/ujpe68.11.717",
    journal = "Ukr. J. Phys.",
    volume = "68",
    number = "11",
    pages = "717",
    year = "2023"
}

@article{reardon_PPTA_2023,
	 author = "Reardon, Daniel J. and others",
    title = "{Search for an Isotropic Gravitational-wave Background with the Parkes Pulsar Timing Array}",
    eprint = "2306.16215",
    archivePrefix = "arXiv",
    primaryClass = "astro-ph.HE",
    doi = "10.3847/2041-8213/acdd02",
    journal = "Astrophys. J. Lett.",
    volume = "951",
    number = "1",
    pages = "L6",
    year = "2023"
}

@article{xu_CPTA_2023,
	author = "Xu, Heng and others",
    title = "{Searching for the Nano-Hertz Stochastic Gravitational Wave Background with the Chinese Pulsar Timing Array Data Release I}",
    eprint = "2306.16216",
    archivePrefix = "arXiv",
    primaryClass = "astro-ph.HE",
    doi = "10.1088/1674-4527/acdfa5",
    journal = "Res. Astron. Astrophys.",
    volume = "23",
    number = "7",
    pages = "075024",
    year = "2023"
}

@article{agazie_nanograv_2023,
	author = "Agazie, Gabriella and others",
    collaboration = "NANOGrav",
    title = "{The NANOGrav 15 yr Data Set: Evidence for a Gravitational-wave Background}",
    eprint = "2306.16213",
    archivePrefix = "arXiv",
    primaryClass = "astro-ph.HE",
    doi = "10.3847/2041-8213/acdac6",
    journal = "Astrophys. J. Lett.",
    volume = "951",
    number = "1",
    pages = "L8",
    year = "2023"
}

@article{agazie_nanograv_2023_noise,
    author = "Agazie, Gabriella and others",
    collaboration = "NANOGrav",
    title = "{The NANOGrav 15 yr Data Set: Detector Characterization and Noise Budget}",
    eprint = "2306.16218",
    archivePrefix = "arXiv",
    primaryClass = "astro-ph.HE",
    doi = "10.3847/2041-8213/acda88",
    journal = "Astrophys. J. Lett.",
    volume = "951",
    number = "1",
    pages = "L10",
    year = "2023"
}

@misc{nanograv_2023_nose_data,
  author       = {Glaser, Joseph and Hazboun, Jeffrey S. and Lam, Michael T. and Lewandowska, Natalia},
  collaboration = "NANOGrav",
  title        = {{Noise Spectra and Stochastic Background Sensitivity Curve for the NG15-year Dataset (1.0.0) [Data set]}},
  month        = jun,
  year         = 2023,
  publisher    = {Zenodo},
  version      = {v1},
  doi          = {10.5281/zenodo.8092346},
  url          = {https://doi.org/10.5281/zenodo.8092346},
}

@article{epta+inpta_2023,
	author = "Antoniadis, J. and others",
    collaboration = "EPTA, InPTA:",
    title = "{The second data release from the European Pulsar Timing Array - III. Search for gravitational wave signals}",
    eprint = "2306.16214",
    archivePrefix = "arXiv",
    primaryClass = "astro-ph.HE",
    doi = "10.1051/0004-6361/202346844",
    journal = "Astron. Astrophys.",
    volume = "678",
    pages = "A50",
    year = "2023"
}

@article{Figueroa_2023,
    author = "Figueroa, Daniel G. and Lizarraga, Joanes and Urio, Ander and Urrestilla, Jon",
    title = "{Strong Backreaction Regime in Axion Inflation}",
    eprint = "2303.17436",
    archivePrefix = "arXiv",
    primaryClass = "astro-ph.CO",
    doi = "10.1103/PhysRevLett.131.151003",
    journal = "Phys. Rev. Lett.",
    volume = "131",
    number = "15",
    pages = "151003",
    year = "2023"
}

@article{caravano_2023,
    author = "Caravano, Angelo and Komatsu, Eiichiro and Lozanov, Kaloian D. and Weller, Jochen",
    title = "{Lattice simulations of axion-U(1) inflation}",
    eprint = "2204.12874",
    archivePrefix = "arXiv",
    primaryClass = "astro-ph.CO",
    doi = "10.1103/PhysRevD.108.043504",
    journal = "Phys. Rev. D",
    volume = "108",
    number = "4",
    pages = "043504",
    year = "2023"
}

@article{durrer_2023,
    author = "Durrer, R. and Sobol, O. and Vilchinskii, S.",
    title = "{Backreaction from gauge fields produced during inflation}",
    eprint = "2303.04583",
    archivePrefix = "arXiv",
    primaryClass = "gr-qc",
    reportNumber = "MS-TP-23-06",
    doi = "10.1103/PhysRevD.108.043540",
    journal = "Phys. Rev. D",
    volume = "108",
    number = "4",
    pages = "043540",
    year = "2023"
}

@article{eckardstein_2023,
    author = "von Eckardstein, Richard and Peloso, Marco and Schmitz, Kai and Sobol, Oleksandr and Sorbo, Lorenzo",
    title = "{Axion inflation in the strong-backreaction regime: decay of the Anber-Sorbo solution}",
    eprint = "2309.04254",
    archivePrefix = "arXiv",
    primaryClass = "hep-ph",
    reportNumber = "ACFI-T23-05, MS-TP-23-38",
    doi = "10.1007/JHEP11(2023)183",
    journal = "JHEP",
    volume = "11",
    pages = "183",
    year = "2023"
}

@article{domcke_2024,
    author = "Domcke, Valerie and Ema, Yohei and Sandner, Stefan",
    title = "{Perturbatively including inhomogeneities in axion inflation}",
    eprint = "2310.09186",
    archivePrefix = "arXiv",
    primaryClass = "astro-ph.CO",
    reportNumber = "IFIC/23-45, FTUV-23-1005.0503, UMN-TH-4226/23, FTPI-MINN-23-18, CERN-TH-2023-186",
    doi = "10.1088/1475-7516/2024/03/019",
    journal = "JCAP",
    volume = "03",
    pages = "019",
    year = "2024"
}

@article{drewes_2024,
    author = "Drewes, Marco and Georis, Yannis and Klasen, Michael and Wiggering, Luca Paolo and Wong, Yvonne Y. Y.",
    title = "{Towards a precision calculation of N $_{eff}$ in the Standard Model. Part III. Improved estimate of NLO contributions to the collision integral}",
    eprint = "2402.18481",
    archivePrefix = "arXiv",
    primaryClass = "hep-ph",
    reportNumber = "CPPC-2024-01, MS-TP-24-06",
    doi = "10.1088/1475-7516/2024/06/032",
    journal = "JCAP",
    volume = "06",
    pages = "032",
    year = "2024"
}

@article{bastero-gil_2024_A,
    author = "Bastero-Gil, Mar and Ferraz, Paulo B. and Ubaldi, Lorenzo and Vega-Morales, Roberto",
    title = "{Schwinger dark matter production}",
    eprint = "2312.15137",
    archivePrefix = "arXiv",
    primaryClass = "hep-ph",
    reportNumber = "UG-FT 329-23, CAFPE 199-23, SISSA 44/2020/FISI, CA21106",
    doi = "10.1088/1475-7516/2024/10/078",
    journal = "JCAP",
    volume = "10",
    pages = "078",
    year = "2024"
}

@article{bastero-gil_2024_B,
    author = "Bastero-Gil, Mar and Ferraz, Paulo B. and Ubaldi, Lorenzo and Vega-Morales, Roberto",
    title = "{Super heavy dark matter from inflationary Schwinger production}",
    eprint = "2311.09475",
    archivePrefix = "arXiv",
    primaryClass = "hep-ph",
    reportNumber = "UG-FT 328-23, CAFPE 198-23, CA21106",
    doi = "10.1103/PhysRevD.110.095019",
    journal = "Phys. Rev. D",
    volume = "110",
    number = "9",
    pages = "095019",
    year = "2024"
}

@article{galloni_2024,
    author = "Galloni, Giacomo and Henrot-Versill{\'e}, Sophie and Tristram, Matthieu",
    title = "{Robust constraints on tensor perturbations from cosmological data: A comparative analysis from Bayesian and frequentist perspectives}",
    eprint = "2405.04455",
    archivePrefix = "arXiv",
    primaryClass = "astro-ph.CO",
    doi = "10.1103/PhysRevD.110.063511",
    journal = "Phys. Rev. D",
    volume = "110",
    number = "6",
    pages = "063511",
    year = "2024"
}

@article{Figueroa_2024,
    author = "Figueroa, Daniel G. and Lizarraga, Joanes and Loayza, Nicol{\'a}s and Urio, Ander and Urrestilla, Jon",
    title = "{Nonlinear dynamics of axion inflation: A detailed lattice study}",
    eprint = "2411.16368",
    archivePrefix = "arXiv",
    primaryClass = "astro-ph.CO",
    doi = "10.1103/PhysRevD.111.063545",
    journal = "Phys. Rev. D",
    volume = "111",
    number = "6",
    pages = "063545",
    year = "2025"
}

@article{miles_meerkat_2024,
	author = "Miles, Matthew T. and others",
    title = "{The MeerKAT Pulsar Timing Array: the first search for gravitational waves with the MeerKAT radio telescope}",
    eprint = "2412.01153",
    archivePrefix = "arXiv",
    primaryClass = "astro-ph.HE",
    doi = "10.1093/mnras/stae2571",
    journal = "Mon. Not. Roy. Astron. Soc.",
    volume = "536",
    number = "2",
    pages = "1489--1500",
    year = "2024"
}

@article{agazie_nanograv_RPL_2025,
	 author = "Agazie, Gabriella and others",
    title = "{The NANOGrav 15 yr Data Set: Running of the Spectral Index}",
    eprint = "2408.10166",
    archivePrefix = "arXiv",
    primaryClass = "astro-ph.HE",
    doi = "10.3847/2041-8213/ad99d3",
    journal = "Astrophys. J. Lett.",
    volume = "978",
    number = "2",
    pages = "L29",
    year = "2025"
}

@article{eckardstein_2025,
    author = "von Eckardstein, Richard and Schmitz, Kai and Sobol, Oleksandr",
    title = "{On the Schwinger effect during axion inflation}",
    eprint = "2408.16538",
    archivePrefix = "arXiv",
    primaryClass = "hep-ph",
    reportNumber = "MS-TP-24-20",
    doi = "10.1007/JHEP02(2025)096",
    journal = "JHEP",
    volume = "02",
    pages = "096",
    year = "2025"
}

@article{sharma_2025,
    author = "Sharma, Ramkishor and Brandenburg, Axel and Subramanian, Kandaswamy and Vikman, Alexander",
    title = "{Lattice simulations of axion-U(1) inflation: gravitational waves, magnetic fields, and scalar statistics}",
    eprint = "2411.04854",
    archivePrefix = "arXiv",
    primaryClass = "astro-ph.CO",
    reportNumber = "NORDITA-2024-040",
    doi = "10.1088/1475-7516/2025/05/079",
    journal = "JCAP",
    volume = "05",
    pages = "079",
    year = "2025"
}

@article{iarygina_2025,
    author = "Iarygina, Oksana and Sfakianakis, Evangelos I. and Brandenburg, Axel",
    title = "{Schwinger effect in axion inflation on a lattice}",
    eprint = "2506.20538",
    archivePrefix = "arXiv",
    primaryClass = "astro-ph.CO",
    reportNumber = "NORDITA-2025-030",
    month = "6",
    year = "2025"
}

@article{Planck:2015zfm,
    author = "Ade, P. A. R. and others",
    collaboration = "Planck",
    title = "{Planck 2015 results. XVII. Constraints on primordial non-Gaussianity}",
    eprint = "1502.01592",
    archivePrefix = "arXiv",
    primaryClass = "astro-ph.CO",
    doi = "10.1051/0004-6361/201525836",
    journal = "Astron. Astrophys.",
    volume = "594",
    pages = "A17",
    year = "2016"
}

@article{Planck:2019kim,
    author = "Akrami, Y. and others",
    collaboration = "Planck",
    title = "{Planck 2018 results. IX. Constraints on primordial non-Gaussianity}",
    eprint = "1905.05697",
    archivePrefix = "arXiv",
    primaryClass = "astro-ph.CO",
    doi = "10.1051/0004-6361/201935891",
    journal = "Astron. Astrophys.",
    volume = "641",
    pages = "A9",
    year = "2020"
}

@article{Bastero-Gil_2025_A,
    author = "Bastero-Gil, Mar and Ferraz, Paulo B. and Torres Manso, Ant{\'o}nio and Ubaldi, Lorenzo and Vega-Morales, Roberto",
    title = "{Schwinger Current in de Sitter Space}",
    eprint = "2503.01981",
    archivePrefix = "arXiv",
    primaryClass = "hep-ph",
    month = "3",
    year = "2025"
}

@article{vonEckardstein_PAI_2025,
    author = "von Eckardstein, Richard and Schmitz, Kai and Sobol, Oleksandr",
    title = "{Gravitational waves from axion inflation in the gradient expansion formalism. Part I. Pure axion inflation}",
    eprint = "2508.00798",
    archivePrefix = "arXiv",
    primaryClass = "astro-ph.CO",
    reportNumber = "MS-TP-25-23",
    doi = "10.1007/JHEP01(2026)018",
    journal = "JHEP",
    volume = "01",
    pages = "018",
    year = "2026"
}

@article{Bastero-Gil_2025_B,
    author = "Bastero-Gil, Mar and Ferraz, Paulo B. and Torres Manso, Ant{\'o}nio and Ubaldi, Lorenzo and Vega-Morales, Roberto",
    title = "{Classical constant electric fields and the Schwinger effect in de Sitter}",
    eprint = "2508.14973",
    archivePrefix = "arXiv",
    primaryClass = "hep-ph",
    month = "8",
    year = "2025"
}

@article{Allahverdi:2010xz,
    author = "Allahverdi, Rouzbeh and Brandenberger, Robert and Cyr-Racine, Francis-Yan and Mazumdar, Anupam",
    title = "{Reheating in Inflationary Cosmology: Theory and Applications}",
    eprint = "1001.2600",
    archivePrefix = "arXiv",
    primaryClass = "hep-th",
    doi = "10.1146/annurev.nucl.012809.104511",
    journal = "Ann. Rev. Nucl. Part. Sci.",
    volume = "60",
    pages = "27--51",
    year = "2010"
}

@article{Kofman:1997yn,
   author = "Kofman, Lev and Linde, Andrei D. and Starobinsky, Alexei A.",
    title = "{Towards the theory of reheating after inflation}",
    eprint = "hep-ph/9704452",
    archivePrefix = "arXiv",
    reportNumber = "IFA-97-28, SU-ITP-97-18",
    doi = "10.1103/PhysRevD.56.3258",
    journal = "Phys. Rev. D",
    volume = "56",
    pages = "3258--3295",
    year = "1997"
}

@article{Amin:2014eta,
    author = "Amin, Mustafa A. and Hertzberg, Mark P. and Kaiser, David I. and Karouby, Johanna",
    title = "{Nonperturbative Dynamics Of Reheating After Inflation: A Review}",
    eprint = "1410.3808",
    archivePrefix = "arXiv",
    primaryClass = "hep-ph",
    doi = "10.1142/S0218271815300037",
    journal = "Int. J. Mod. Phys. D",
    volume = "24",
    pages = "1530003",
    year = "2014"
}

@dataset{LIGOScientificCollaboration2021GWTC3,
  author       = {{LIGO Scientific Collaboration} and {Virgo Collaboration} and {KAGRA Collaboration}},
  title        = {{GWTC-3: Compact Binary Coalescences Observed by LIGO and Virgo During the Second Part of the Third Observing Run — Data behind the figures}},
  year         = {2021},
  publisher    = {Zenodo},
  doi          = {10.5281/zenodo.5571767},
  url          = {https://doi.org/10.5281/zenodo.5571767},
  note         = {[Data set]}
}

@article{Thrane:2013oya,
    author = "Thrane, Eric and Romano, Joseph D.",
    title = "{Sensitivity curves for searches for gravitational-wave backgrounds}",
    eprint = "1310.5300",
    archivePrefix = "arXiv",
    primaryClass = "astro-ph.IM",
    doi = "10.1103/PhysRevD.88.124032",
    journal = "Phys. Rev. D",
    volume = "88",
    number = "12",
    pages = "124032",
    year = "2013"
}

@article{PencilCode:2020eyn,
    author = "Brandenburg, A. and others",
    collaboration = "Pencil Code",
    title = "{The Pencil Code, a modular MPI code for partial differential equations and particles: multipurpose and multiuser-maintained}",
    eprint = "2009.08231",
    archivePrefix = "arXiv",
    primaryClass = "astro-ph.IM",
    reportNumber = "NORDITA-2020-087",
    doi = "10.21105/joss.02807",
    journal = "J. Open Source Softw.",
    volume = "6",
    number = "58",
    pages = "2807",
    year = "2021"
}

@article{Ferreira:2017lnd,
    author = "Ferreira, Ricardo Z. and Notari, Alessio",
    title = "{Thermalized Axion Inflation}",
    eprint = "1706.00373",
    archivePrefix = "arXiv",
    primaryClass = "astro-ph.CO",
    doi = "10.1088/1475-7516/2017/09/007",
    journal = "JCAP",
    volume = "09",
    pages = "007",
    year = "2017"
}

@article{vonEckardstein:2025jug,
        author        = "von Eckardstein, Richard",
        title         = "GEFF: The Gradient Expansion Formalism Factory -- A tool for inflationary gauge-field production",
        eprint        = "2510.12644",
        archivePrefix = "arXiv",
        primaryClass  = "astro-ph.CO",
        year          = "2025"}




\end{document}